\documentclass[hidelinks,journal]{IEEEtran}
\IEEEoverridecommandlockouts

\usepackage[square, numbers, comma, sort&compress]{natbib} 

\usepackage{amsmath,amssymb,amsfonts, bm}
\usepackage{graphicx}
\usepackage{subcaption}
\usepackage{algorithm,algcompatible}
\usepackage{booktabs}
\usepackage{textcomp}
\usepackage{epstopdf}
\usepackage{changepage}   
\usepackage{balance}
\usepackage[export]{adjustbox}

\newtheorem{assumption}{\bf Assumption}
\newtheorem{theorem}{\bf Theorem}
\newtheorem{remark}{\bf Remark}

\newenvironment{proof}{{\it {Proof:}}}{\hfill $\blacksquare$}

\DeclareMathOperator{\bR}{\mathbb R}
\DeclareMathOperator{\bE}{\mathbb E}

\DeclareMathOperator{\tp}{{\mathsf{T}}}
\DeclareMathOperator{\col}{\mathrm{col}}

\usepackage{color}
\definecolor{mycolour1}{rgb}{0.0157, 0.1843, 0.4}
\definecolor{mycolour2}{rgb}{0.2667, 0.7372, 0.8471}

\begin{document}

	\title{
		Concurrent Active Learning in Autonomous Airborne Source Search: Dual Control for Exploration and Exploitation
		\thanks{This work was supported by the UK Engineering and Physical Sciences Research Council (EPSRC) Established Career Fellowship ``Goal-Oriented Control Systems: Disturbance, Uncertainty and Constraints" under the grant number EP/T005734/1.}\thanks{The authors are with the Department of Aeronautical and Automotive Engineering, Loughborough University, Loughborough,  LE11 3TU, U.K. (emails: z.z.li@lboro.ac.uk; w.chen@lboro.ac.uk;  j.yang3@lboro.ac.uk).}\author{\IEEEauthorblockN{Zhongguo Li, {\it Member, IEEE}, Wen-Hua Chen, {\it Fellow, IEEE}, and Jun Yang, {\it Senior Member, IEEE}}}
	}
	
	\maketitle
	
	\begin{abstract}
		In this paper, a concurrent learning framework is developed for source search in an unknown environment using autonomous platforms equipped with onboard sensors. Distinct from the existing solutions that require significant computational power for Bayesian estimation and path planning, the proposed solution is computationally affordable for onboard processors. A new concept of concurrent learning using multiple parallel estimators is proposed to learn the operational environment and quantify estimation uncertainty. The search agent is empowered with dual capability of exploiting current estimated parameters to track the source and probing the environment to reduce the impacts of uncertainty, namely Concurrent Learning based Dual Control for Exploration and Exploitation (CL-DCEE). In this setting, the control action not only minimises the tracking error between future agent's position and estimated source location, but also the uncertainty of predicted estimation. More importantly, the rigorous proven properties such as the convergence of CL-DCEE algorithm are established under mild assumptions on noises, and the impact of noises on the search performance is examined. Simulation results are provided to validate the effectiveness of the proposed CL-DCEE algorithm. Compared with the information-theoretic approach, CL-DCEE not only guarantees convergence, but produces better search performance and consumes much less computational time. 
	\end{abstract}
	

	\section{Introduction}
	
	\IEEEPARstart{P}{ublic} concern towards safety and health issues has been significantly exacerbated during the past few decades, due to rapid industrial development and increasing risks of terrorism. Identifying sources of airborne release (including chemical, biological, radiological and nuclear (CBRN) materials) is one of the most important tasks in disaster management and environment protection \cite{Hutchinson2017review}. In the early literature, source term estimation (STE) is mainly supported by onsite measurement using static sensor networks that are deployed beforehand in some specific areas of potential risks \cite{Hutchinson2017review, Lennart1998source, singh2015inverse, Rao2007source}. This type of strategy is very costly, and only feasible for high-risk industry, e.g. nuclear power plants \cite{tsitsimpelis2019review}. 
	Recently, significant research efforts have been dedicated to the development of dynamic estimation of airborne release assisted by mobile platforms, for example, autonomous ground robots \cite{Hutchinson2018TCST,Chen2021DCEE, vergassola2007nature, pang2006chemical} and unmanned aerial vehicles (UAVs) \cite{Hutchinson2019magazine, Hutchinson2019fiedrobotics, yang2020optimal}. Compared with conventional static methods, autonomous search is much more flexible and cost-effective in emergent accident management. Comprehensive surveys of recent progress on source localisation using autonomous vehicles have been reported in several review papers \cite{Hutchinson2017review, chen2019odor,villa2016overview}.

	There are various methods dealing with this problem \cite{Hutchinson2017review,Chen2021DCEE}. Among them, informative path planning (IPP) becomes increasingly popular, e.g. Infotaxis \cite{vergassola2007nature} and Entrotaxis \cite{Hutchinson2018Entrotaxis}. Vergassola \emph{et al.} \cite{vergassola2007nature} proposed an informative search approach, referred as Infotaxis, by which the agent moves to the next position that is expected to minimise uncertainties of the posterior distribution. Hutchinson \emph{et al.} \cite{Hutchinson2018Entrotaxis} developed Entrotaxis algorithm that steers the agent to search over the most uncertain area in the next movement. More recently, some advanced versions of the above algorithms have been developed aiming to improve their robustness, search speed and accuracy in more complex search environment, including  Infotaxis II \cite{Branko2016study}, Entrotaxis-jump \cite{zhao2020entrotaxis-jump} and hybrid of Infotaxis and Entrotaxis \cite{zhao2020searching}. Zhao \emph{et al.} \cite{zhao2020entrotaxis-jump} proposed an Entrotaxis-jump algorithm for source search in large-scale road networks with intermittent jumps to traverse obstacles \cite{zhao2020entrotaxis-jump}. In \cite{zhao2020searching}, a number of hybrid strategies combining Infotaxis and Entrotaxis are proposed in the random obstructed environment with forbidden search areas. Essentially, information-theoretic approaches aim to reduce uncertainties of estimated source location and unknown environment parameters. Therefore, the reward function,  which the agent targets to maximise while making its next movement, is defined according to the information gain using some informative measures \cite{sun2011planning}, for example, entropy, Kullback-Leibler divergence and variance.

	Apart from information-theoretic methods, another main branch for source seeking is the optimisation approach. Stochastic extremum seeking is employed to direct a nonholonomic unicycle towards the maximum of an unknown signal field \cite{liu2010stochastic}. A simultaneous perturbation stochastic approximation approach (SPSA) is developed to approach a minimiser of a source signal for environment with and without obstacles \cite{Ramirez-Llanos2018stochastic}, which can be traced back to the early work in \cite{sapll1992multivariate}. An adaptive gradient climbing method is designed for cooperative mobile sensors to seek the optimiser of environmental field in \cite{ogren2004cooperative}. It is worth mentioning that convergence guarantees of those algorithms have been well-studied by leveraging advanced control and stochastic approximation techniques. In those learning based control approaches, the unknown parameters of the environment are \emph{passively} updated.
	In Bayesian framework developed in \cite{Bourgault2006optimal}, optimal trajectory planning uses the current estimate to plan the future trajectory in order to maximise cumulative probability of detection or minimise mean time of detection.
	In fact, the control actions of aforementioned works are designed based on the principle of certainty equivalence by treating the current estimation of the environment as the true parameters. As a result, estimation uncertainty is not explicitly taken into account when the search agent makes next movement.

	Recently, Chen {\it et al.} \cite{Chen2021DCEE} have reformulated the autonomous search problem from a control-theoretic perspective, referred as Dual Control for Exploration and Exploitation (DCEE). The ultimate goal of autonomous search is to design a control strategy that can navigate the agent to an \emph{unknown release} in an \emph{unknown environment}, which is a well-posed goal-oriented control problem.  Distinct from traditional control settings where operational systems are manipulated by following predefined references or setpoints, the autonomous search problem does not have such a given reference or path that can directly lead the agent to the source. Instead, the agent is required to \emph{explore} the operational environment to learn the source parameters, and at the same time \emph{exploit} its belief to move towards the source. From the perspective of dual control, information-theoretic approach is a pure exploration strategy aiming at collecting information from unknown environment while model predictive control (MPC) is a pure exploitation approach targeting at making full use of current uncertain estimation. 
	This novel dual control framework achieves a natural balance between the two objectives, and has demonstrated superior performance in real experiments compared with MPC and IPP.
	
	Although DCEE offers a conceptually promising framework in autonomous search, similar to all the existing information-theoretic approaches, currently it still suffers from two drawbacks: computational burden and no rigorous analysis of its properties such as stability and convergence.  IPP and DCEE approaches demand massive computational burden imposed by nonlinear particle filters and optimisation based path planning \cite{li2011odor, Hutchinson2018Entrotaxis, ristic2003beyond}. More specifically, the Bayesian inference engine is involved in the optimisation loop for informative path planning since the influence of the control action on the predicted posterior of the estimated source and environment parameters is evaluated at each iteration. Consequently, the search algorithm is restricted to be a set of certain moving directions with a fixed step size, and currently carried out remotely in a control centre, rather than onboard on the mobile robot in the existing experimental studies \cite{Hutchinson2018TCST, Chen2021DCEE, Hutchinson2019magazine, Rhodes2021autonomous}. This significantly restricts their practical applications in autonomous search in a wide area or an extreme environment. Hence, less computationally demanding alternatives are needed so that online computation can be achieved by portable processors on mobile platforms. To the best of our knowledge, currently there are quite limited proven properties of IPP autonomous search strategies. This was actually a main motivation of formulating it from a control-theoretic approach in \cite{Chen2021DCEE} since it enables to get access to rich knowledge and tools available in control theory.  No rigorous analysis on convergence is provided but mainly illustrated through extensive simulation and experimental studies. In IPP and DCEE, the agent's movement (path planning) and source estimation (environment acquisition) are strongly coupled: the agent takes actions according to the current estimation of source parameters and the estimators update their knowledge by using the concentration collected at agent's new position determined by path planning. This coupling, together with noisy measurement, environment turbulence, complicated particle filtering and optimisation involved in implementation of the search strategy, makes the rigorous analysis of theoretic properties of these search strategies quite challenging. Those two critical limitations motivate this study. 
	
	In this paper, inspired by the concept of DCEE, we propose a concurrent learning based DCEE scheme with multiple estimators that encompasses dual effects: driving the agent to the believed location by exploiting current estimation, and reducing uncertainties by exploring the unknown operational environment, which is referred as Concurrent Learning based Dual Control for Exploration and Exploitation (CL-DCEE) for the sake of simplicity. The underlying principle of the concurrent learning scheme advocated in this paper is distinct from the classic dual control in handling the two intricate coupling elements of the \emph{system} and the \emph{environment}. Existing dual control approaches impose a probing effect on the \emph{system} itself, for example, state estimation in stochastic control \cite{mesbah2018stochastic, chen2018approximating} and parameter estimation in adaptive control \cite{bugeja2009dual, filatov2000survey}. On the other hand, the dual effect introduced in our formulation is used to explore the operational \emph{environment}, as our objective is to acquire a better understanding of the unknown environment such that the agent is able to approach the true source location. 
	
	To address the two challenges of computational burden and proven properties, two approaches are proposed in this paper. Instead of implementing computationally demanding particle filtering, an efficient multi-estimator scheme is proposed for source and environment learning. The number of estimators used in CL-DCEE is much smaller than that of particles required for Bayesian filters in information-theoretic algorithms. These estimators run in parallel from a set of randomly started initial estimates. 
	There are several fundamental incentives promoting us to employ multiple concurrent estimators.  
	First, compared with employing a single estimator, this multi-estimator approach provides a means to quantify uncertainty associated with source estimators, which is of great importance to empower the search agent with \emph{dual capability of exploration and exploitation}. Secondly, it significantly improves performance and robustness over a single estimator. The performance of a single estimator (such as an observer or learning machine) is often severely influenced by the initialisation and setting of the individual estimator, for example, state estimation \cite{li2011generalized}, disturbance observer \cite{chen2015disturbance} and parameter adaptation \cite{goodwin1987parameter}. 
	To our knowledge, there is few result on multi-estimator assisted control algorithms. Devising multiple parallel estimators for the source parameters is conducive to eliminating undesirable behaviour caused by random initialisation of an individual, and also it allows us to take advantage of {\it a priori} probability density function (PDF) of source parameters.
	To further reduce the computational load, effective gradient-based optimisation algorithms are utilised to replace the complicated path planning process in the existing methods. It is shown that by combining these two techniques, we are able to reduce the computational load by 100 times while significantly increasing the admissible control set. Most importantly, we establish theoretical guarantee of convergence of the CL-DCEE algorithm, and analyse the steady-state performance of the estimation and search by directly linking them with sensor and environment characters.

	In summary, the key contributions of this paper are enumerated  as follows. 
	\begin{enumerate}
		\item A concurrent active learning algorithm with multiple environment estimators is developed, which achieves a balanced trade-off between \emph{exploitation} of believed source location and \emph{exploration} of uncertain environment, that is, simultaneously navigating the agent to the source and reducing the impacts of uncertainties associated with the acquired environment knowledge.
		\item The convergence of the proposed autonomous search algorithm, CL-DCEE, is rigorously established under sensor and control noises, by leveraging a memory based stochastic approximation and a gradient based path planning strategy.  
		\item The proposed CL-DCEE provides a \emph{computationally efficient} solution for autonomous search of airborne source release. Simulation and experiment results are provided to demonstrate the performance of the proposed method in comparison with information-theoretic approaches. Our solution shows superior performance with significant reduction in computational time.   
	\end{enumerate}

	The remainder of this paper is organised as follows. In Section~\ref{sec: 2}, we formulate the autonomous search problem and develop feasible value functions for the path planning and estimation. In Section~\ref{sec: 3}, CL-DCEE algorithm is proposed by deploying multiple environment estimators. Section~\ref{sec: 4} provides simulation results and detailed discussions in comparison with existing approaches. Experimental study using a real dataset is presented in Section~\ref{sec: 5}. Section~\ref{sec: 6} concludes this paper.

	\section{Problem Formulation}\label{sec: 2}

	\subsection{Agent Modelling}
	The searching agent is considered as a fully autonomous vehicle, for example, a mobile robot or a UAV, which is equipped with chemical/biological sensors. We assume that the agent has been devised with a low-level controller that can steer the agent to the desired position directed by high-level decision-making process. Therefore, the dynamics of the search agent can be simplified as 
	\begin{equation}
	    \bm p_{k+1} = \bm p_k + \bm u_k + w_k
	\end{equation}
	where $ \bm p_k =[p_{k,x}, p_{k,y},p_{k,z}]^{\tp} \in \Omega \subseteq \bR^3 $ denotes the position of the searching agent at current step $ k $ with $ \Omega $ being {\color{black}a convex and compact searching space}, $ \bm u_k \in \mathcal U \subseteq \bR^3 $ is the control action with $ \mathcal U $ being the admissible set of actions, and $w_k$ is the control error. It is worth mentioning that in this paper the admissible set $ \mathcal U $ can be continuous, which is distinct from the existing results in \cite{Hutchinson2018TCST,Hutchinson2019magazine, Chen2021DCEE} where the search is restricted to certain directions with a fixed step size.

	\subsection{Dispersion Modelling}
	Atmospheric transport and dispersion model (ATDM), governing the spatial-temporal diffusion of the pollutant materials, is utilised to predict the concentration in space and time, given the parameters of a release. In this paper, we denote  true source parameters as $ \bm\Theta_s = [\bm  s^{\tp}, q]^{\tp} \in \bR^4 $ with $ \bm s=[s_x,s_y,s_z ]^{\tp}  \in\bR^3 $ being the position of the source and $q\in \bR^+ $ representing a positive release rate. The dispersion model is given by
	\begin{equation}\label{eqn: ATD model}
			\begin{aligned}
				\mathcal{M}\left(\bm {p}_k, \bm \Theta_s\right) = & \frac{q}{4 \pi \zeta_{s 1}|| \bm{p}_k-\bm s \|} \exp \left[\frac{-\left\|\bm{p}_k-\bm s\right\|}{\zeta}\right]  \\ & \times
				\exp \left[\frac{-\left(p_{k,x}-s_{x}\right) u_{s} \cos \rho_{s}}{2 \zeta_{s 1}}\right] 
				\\ & \times \exp \left[\frac{-\left(p_{k,y}-s_{y}\right) u_{s} \sin \rho_{s}}{2 \zeta_{s 1}}\right]
			\end{aligned}
	\end{equation}
	where  environmental parameters include the wind speed $u_s$, wind direction $\rho_{s}$, diffusivity $\zeta_{s1}$, the particle lifetime $\zeta_{s2}$, and a composite coefficient $\zeta=\sqrt{\frac{\zeta_{s 1} \zeta_{s 2}}{1+\left(u_{s}^{2} \zeta_{s 2}\right) /\left(4 \zeta_{s 1}\right)}}$.

	\subsection{Sensor Modelling}
	Information collection in autonomous search of an airborne release is mainly from onboard chemical/biological sensors. As the search agent moves to a new position, concentration measurement will be taken. The agent is required to remain at current position for a short period to obtain a reliable reading, referred as the sampling time. 
	It is likely that the sensors fail to detect any meaningful readings. Sensors deployed on mobile platforms are usually low-price portable devices whose performances are seriously constrained by limited power supply and local turbulence. As summarised in~\cite{Hutchinson2019fiedrobotics}, there are several scenarios that may cause a non-detection event: 1) there is no chemical material present in the current position, or present but out of the detection range of the sensor; 2) the source is present in the position but the detector fails to receive any material due to intermittent turbulence; 3) the concentration is below a pre-specified threshold, and is thus classified as a non-detection event.

	In summary, the sensor reading can be modelled as 
	\begin{equation}\label{eqn: sensor model}
	    z(\bm{p}_{k} )=\left\{\begin{array}{ll}
	    \mathcal{M}\left(\bm{p}_{k}, \bm \Theta_s \right)+v_{k}, \ \  & D =1 \\
	    \bar{v}_{k}, \ \ & D =0
	    \end{array}\right.
	\end{equation}
	where $\mathcal{M}$ is the true chemical concentration, $ D $ denotes either a detection event $ D=1 $ or a non-detection event $D=0$, and $ v_k $ and $\bar v_k $ represent additive Gaussian noises imposed on the sensor readings. The probability distribution of $ D $ is a random process, usually characterised by Poission distribution. 
	For theoretical analysis, we assume that the sensor can always receive concentration with additive white noises. In simulation and experiment, we will test the performance of the proposed algorithm with and without sensor dropouts.

	\subsection{Objective Function Construction}

	The dispersion model in (\ref{eqn: ATD model}) is referred as an isotropic plume (IP) model \cite{vergassola2007nature}. There are many other commonly-used dispersion models, such as Gaussian plume (GP) and computational fluid dynamics (CFD) \cite{holmes2006review}. Nevertheless, the model can be understood as a concentration function that possesses the highest value at the release centre and decreases monotonically as the increase of the distance to the centre (in terms of expectation). Thus, it can be used to formulate an optimisation objective for the autonomous agent, by taking the position $ \bm p_k $ as the optimisation variable. Following the convention in optimisation theory, the objective function is defined as 
	\begin{equation}\label{eqn: concentration optimisation}
	g\left(\bm {p}_k, \bm \Theta_k\right) = (\mathcal M_0-\mathcal{M}\left(\bm {p}_k, \bm \Theta_k\right)  )^2
	\end{equation}
	where $ \mathcal M_0 $ is a predefined upper bound of the concentration measurement. It is clear that the optimal solution of (\ref{eqn: concentration optimisation}) is $ \bm p_k^* = \bm s $, where $\mathcal{M}\left(\bm {p}^*_k, \bm \Theta_s\right) $ is maximised.

	To estimate the source location $\bm s$ and release rate $ q $ based on available measurements, we may define an additional value function taking the source term as the decision variable. Least square methods can serve for this purpose, given by 
	\begin{equation}\label{eqn: value function}
		 f(\bm \Theta_i, \bm p_i) =   \left[\mathcal M(\bm p_i, \bm\Theta_i) - z(\bm p_i)\right]^2
	\end{equation}
	where  $ z(\bm p_i) $ denotes the measured concentration at agent position $ \bm p_i , \forall i = 1, \dots, k $.

	\section{Concurrent Learning for Dual Control with Exploration and Exploitation}\label{sec: 3}

	\subsection{Framework of Concurrent Learning with Dual Effects}
	Suppose that the search agent aims to minimise the objective function defined in (\ref{eqn: concentration optimisation}) at each step. Then, it will move to the position where the maximum concentration measurement is expected based on the current belief of the source, i.e. its estimate $\bm \Theta_k$. This type of method can be understood as a pure exploitation strategy, which aims to make full use of the current estimation of the source parameters $ {\bm \Theta}_k$. If the source estimators were perfect without any noisy disturbances or uncertainties, then exploiting the trustworthy estimator would accelerate the speed of finding the source and its associated parameters. Nonetheless, the operational environment is often uncertain, and perfect estimation is never available. This motivates the development of active learning methods to autonomously explore the unknown environment so as to construct more accurate estimators.

	The dual control framework for autonomous source search and estimation was firstly introduced by Chen \emph{et al.} \cite{Chen2021DCEE} recently. The goal is to drive the agent towards the believed position of a release and in the meanwhile reduce uncertainty associated with the estimation of the target position. Generally speaking, uncertainty is often measured in a stochastic sense for probability density function of a variable. In \cite{Hutchinson2018Entrotaxis}, particle filters are utilised to estimate the source parameters and the uncertainty associated with the estimated source target. However, it requires a large number of particles to support the  Bayesian inference engine, which incurs heavy computational burden. Quantifying the uncertainty is of great importance, as demonstrated in our previous works \cite{Chen2021DCEE, Hutchinson2018Entrotaxis, Hutchinson2018TCST, Hutchinson2019fiedrobotics}. To alleviate this problem, we thus introduce a set of $N$ source term estimators and they are initialised according to the prior knowledge of the source parameters. It is worth emphasising that the number of estimators $N$ is much smaller than that of particles in Bayesian filters as shown later.

	The concentration information collected up to time step $ k $ is denoted by $\bm{\mathcal{Z}}_{k}:=\left\{z\left(\bm {p}_{1}\right), z\left(\bm{p}_{2}\right), \ldots, z\left(\bm{p}_{k}\right)\right\}$. Let $ \bm{\Theta}_{k}^i $ be the source term estimation of the $ i $th estimator at the $ k $th measurement, and $ \bar{\bm \Theta}_k := \frac 1N \sum_{i=1}^N \bm{\Theta}_{k}^i $ as the nominal estimation, i.e. the mean, of the source parameters. The posterior distribution of source estimation can be represented by $\rho_{k | k}:=p\left(\bm \Theta | \bm{\mathcal{Z}}_{k}\right)$ at time $ k $. When the search agent moves to a new position directed by the control input $ \bm u_k $, the hypothetical posterior distribution of source estimation will be updated as $\hat{\rho}_{k+1 | k}:=p\left(\bm \Theta | \bm{\mathcal Z}_{k+1|k} \right)$ where $ \bm{\mathcal Z}_{k+1|k} = \{ \bm{\mathcal Z}_{k}, \hat z_{k+1|k} \} $, and consequently the future belief of concentration can be regarded as a random variable conditional on $\bm u_k$, denoted as $\hat{z}_{k+1|k} \sim p\left(\hat{z}_{k+1|k} | \bm {u}_{k}\right)$. As a result, the control input $ \bm u_k $ will not only affect the future concentration measurement at agent's new position but also affect the belief of future measurement distribution.

	Motivated by the above discussion, the control input $ \bm u_k $ should be designed to navigate the agent to the position where the \emph{predicted} posterior of the measurement $\hat{z}_{k+1|k}$ is close to the predefined threshold $\mathcal M_0 $. Therefore, the conditional cost function can be formulated as 
	\begin{subequations}\label{eqn: dual objective} \begin{align}
		\min_{\bm u_k \in \mathcal U} J(\bm {u}_k) = &
		\min _{\mathbf{u}_{k} \in \mathcal U} \mathbb{E}_{\bm \Theta}\left[\mathbb{E}_{\hat{z}_{k+1|k}}\left[\left(\mathcal M_0 - \hat{z}_{k+1|k}  \right)^{2} | \bm{\mathcal{Z}}_{k+1 | k}\right]\right] \\
		\text{subject to} \ \ & \bm p_{k+1|k} = \bm p_k + \bm u_k +w_k .
		\end{align}
	\end{subequations}
	
	The physical interpretation is, based on all the available information including priors and available measurements, we would like the robot moving to a place where the \emph{predicted} maximum concentration is located. This mechanism is behind Chemotaxis, a widely adopted search strategy in nature from bacteria to human being \cite{stock2009chemotaxis}. We show that the control action $ \bm u_k $, obtained from the optimisation problem in (\ref{eqn: dual objective}), \emph{implicitly} carries dual effects. We define $\bar z_{k+1|k}$ as the nominal predicted concentration of the future virtual measurements, i.e. the mean of $p\left(\hat{z}_{k+1|k} | \bm {u}_{k}\right)$, written as 
	\begin{equation}
	\bar z_{k+1|k} := \bE\left[\hat{z}_{k+1|k} |  \bm{ \mathcal{Z}}_{k+1 | k} \right] .
	\end{equation}
	Based on the definition of $\bar z_{k+1|k}$, we can further define $ \tilde z_{k+1|k} = \hat z_{k+1|k}   - \bar{ z}_{k+1|k}$. Therefore, the objective function can be reformulated as 
	\begin{equation}\label{eqn 8}
	J(\bm {u}_k) =
	\mathbb{E}_{\bm \Theta, \hat{z}_{k+1|k}}\left[\left(\mathcal M_0 - \bar{ z}_{k+1|k} - \tilde z_{k+1|k}  \right)^{2} | \bm{ \mathcal{Z}}_{k+1 | k}\right].
	\end{equation}
	Expanding (\ref{eqn 8}) leads to 
	\begin{equation}\label{eqn 9}
	\begin{aligned}
	J\left(\mathbf{u}_{k}\right)=& \mathbb{E}\left[\left( \mathcal M_0- \bar{ z}_{k+1|k}   \right)^{2} | \bm{\mathcal{Z}}_{k+1 | k}\right] \\
	&-2 \mathbb{E}\left[ \tilde z_{k+1|k}\left(\mathcal M_0- \bar{ z}_{k+1|k}   \right) | \bm{\mathcal{Z}}_{k+1 | k}\right] \\
	&+\mathbb{E}\left[\tilde z_{k+1|k}^2 | \bm{\mathcal{Z}}_{k+1 | k}\right].
	\end{aligned}
	\end{equation}
	Since $\mathcal M_0 $ and $ \bar{ z}_{k+1|k}$ are deterministic variables and $ \mathbb{E}\left[ \tilde z_{k+1|k}\right] = 0$, it follows that 
	\begin{equation}\label{eqn: reformulated dual objective}
	\begin{aligned}
	J\left(\mathbf{u}_{k}\right)= \mathbb{E}\left[\left( \mathcal M_0- \bar{ z}_{k+1|k}   \right)^{2} | \bm{\mathcal{Z}}_{k+1 | k}\right] 
	+\mathbb{E}\left[\tilde z_{k+1|k}^2 | \bm{\mathcal{Z}}_{k+1 | k}\right]
	\end{aligned}.
	\end{equation}
	
	In case of $ N $ estimators, we have $ \bar{ z}_{k+1|k} =  \frac 1N \sum_{i=1}^N   \hat z_{k+1|k}^i  $, with $ \hat z_{k+1|k}^i $ being the predicted measurement at agent's future position $\bm p_{k+1|k}$ based on the $ i $th source estimator $\bm \Theta_k^i$. Then, the optimisation problem for CL-DCEE can be formulated as 
	\begin{subequations}\label{eqn: CL-DCEE formulation} \begin{align}
		&  \min_{\bm u_k \in \mathcal U} J(\bm {u}_k) = \min_{\bm u_k \in \mathcal U}  \ \left [ \left (\mathcal M_0 - \bar{ z}_{k+1|k} \right)^2  + {\mathcal P}_{k+1| k} \right]  \label{eqn: N dual objective}   \\
		&\  \mathcal P_{k+1| k} := \frac 1N \sum_{i=1}^N  (\hat{ z}_{k+1|k}^i - \bar{ z}_{k+1|k} )^{2}  \label{eqn: predicted variance} \\
		&	\  \bm p_{k+1|k} = \bm p_k + \bm u_k +w_k.
		\end{align}
	\end{subequations}

	\begin{remark}
		According to the definition of $ \mathcal P_{k+1| k} $ in (\ref{eqn: predicted variance}), it is clear that $ \mathcal P_{k+1| k} $ is the \emph{predicted} variance of $\hat{ z}_{k+1|k}^i, \forall i =1,\dots, N, $ given that each estimator has a uniform weight of $ 1/N $. The value function in (\ref{eqn: N dual objective}) consists of two parts: the first part exploits current information by navigating the agent towards the believed position of higher concentration, and the second part aims to gather more information by reducing the variance of future virtual measurements. In short, \emph{exploitation} drives the agent to the position where high concentration is expected by minimising $ (\mathcal M_0 - \bar{ z}_{k+1|k})^2  $, while the \emph{exploration} effect makes the agent search over some position that can reduce the predicted variance of the concentration by minimising ${\mathcal P}_{k+1| k} $. 
		The first term in (\ref{eqn: N dual objective}) is closely related to the exploitation objective \eqref{eqn: concentration optimisation} by navigating the agent to higher concentration field. Because the estimators are updated according to the least square function in \eqref{eqn: value function}, the second term in  (\ref{eqn: N dual objective}) is determined by the updating mechanism of the estimators.
	\end{remark}

	\begin{remark}
		Recently, how to balance between exploration and exploitation has aroused extensive discussions and arguments in many areas, in particular artificial intelligence, optimisation and decision-making \cite{mesbah2018stochastic, Chen2021DCEE}. In some cases, artificial weights are introduced on purpose to impose both effects \cite{mesbah2018stochastic}. From the above formulation process of our framework, a \emph{natural} balance between the two effects is derived from a physically meaningful cost function. Accordingly, our framework eliminates the requirement for choosing trade-off weights.
	\end{remark}

	To obtain the variance ${\mathcal P}_{k+1| k} $, we resort to the classical principle of predicting variance estimation in extended Kalman filters~\cite{welch1995introduction}, which can be formulated as 
	\begin{equation}\begin{aligned}\label{eqn: 10}
	& \mathcal P_{k+1|k} = { \mathcal P}_{k|k} \bm F_{k+1}^{\tp} \bm F_{k+1} \\ 
	&  F_{k+1}^i =  \frac{\partial   \mathcal{M}\left(\bm {p}_{k},  {\bm\Theta}_k^i \right) }{\partial \bm p} 
	\end{aligned}
	\end{equation} 
	where $ {\mathcal P}_{k| k} = \frac 1N \sum_{i=1}^N  ( z_k^i - \bar{ z}_{k} )^{2} $ denotes current variance of estimated measurement,  $z_k^i = \mathcal M \left(\bm {p}_{k},  {\bm\Theta}_k^i \right) $, $ \bar{ z}_{k} =  \frac 1N \sum_{i=1}^N   z_{k}^i  $ and $ \bm F_{k+1} = \col[F_{k+1}^1,\dots, F_{k+1}^N]$. In (\ref{eqn: 10}), $ \bm p_{k+1|k} $ is future position of the agent, which will serve as the optimisation variable in the CL-DCEE algorithm.

	Now, we can present the gradient-based optimisation algorithm for the source term estimation and path planning. For notational convenience, we will use 
	\begin{equation}
	    y(\bm {p}_{k}, \bm \Theta_k) =(\mathcal M_0 - \bar{ z}_{k+1|k})^2
	\end{equation}  
to denote the first term in the dual objective (\ref{eqn: N dual objective}). Inspired by the memory based regression parameter estimation methods \cite{ortega2020modified}, the $N$ source estimators can be updated according to 
		\begin{equation}\label{alg: N source term gradient update}
        	\begin{aligned}
				\bm \Theta_{k+1}^i & = \bm \Theta_{k}^i -  \sum_{t = k-q+1}^{k} \eta_t \tilde \nabla_{\bm \Theta}  f(\bm \Theta_{k}^i, \bm p_t)   , \  \forall i=1,2,\dots, N
			\end{aligned}
	\end{equation}
where $q$ is a positive integer denoting the number of past measurement used at the $k$th iteration.
	The path planning is given by 
		\begin{equation}\label{alg: agent movement with dual obective}
		\begin{aligned}
			& \bm p_{k+1} = \bm p_k + \bm u_k + w_k \\
			& \bm u_k =  - \delta_k \left[  \nabla_{\bm p}  y \left(\bm {p}_{k}, \bm \Theta_k \right)   +  \nabla_{\bm p} \mathcal P_{k+1|k}   \right] 
		\end{aligned}
	\end{equation}
	where $ \eta_t, \delta_k$ are constant step sizes to be designed, and $\bm \Theta_k$ represents the collection of all $N$ estimators.
	Note that $ \nabla_{\bm p}  y\left(\bm {p}_{k},\bm \Theta_k\right) $ and $\nabla_{\bm p} \mathcal P_{k+1|k}$ are pure predictions without measurement noises.   Basically, algorithms (\ref{alg: N source term gradient update}) and (\ref{alg: agent movement with dual obective}) use gradient descent method to ensure that the agent moves towards the believed position of a release and the source estimators converge to the true parameters that minimise the least square function in (\ref{eqn: value function}).

	The approximated gradients of the least square function (\ref{eqn: value function}) can be written as
	\begin{equation}\label{eqn: approximated gradient}\begin{aligned}
	\tilde \nabla_{\bm \Theta} f(\bm \Theta_k^i,\bm {p}_{k}) 
	=  \nabla_{\bm \Theta} f(\bm \Theta_k^i,\bm {p}_{k}) + \mu_k \\
	\end{aligned}
	\end{equation}
	where $\mu_k$ denotes the gradient noises, which can be regarded as a source of perturbation to the true gradient caused by the sensory noises.
	
	For convergence analysis, some basic assumptions on the gradient and control noises are introduced in the following. 
	\begin{assumption}\label{asm: sensor noise}
		The noise $\mu_k$ in \eqref{eqn: approximated gradient} satisfies the following properties:
		\begin{align}
			\mathbb{E}\left[\mu_k \right] &=0    \label{eqn: sensor noise 1} \\
			\mathbb{E}\left[\left\| \mu_k \right\|^{2} \right] & \leq  
			{\varrho}^{2} \label{eqn: sensor noise 2} 
		\end{align}
		where $\varrho $ is a positive constant. The position control error has similar properties:
		\begin{align}
			\mathbb{E}\left[w_k \right] &=0  \label{eqn: position noise 3}  \\
			\mathbb{E}\left[\left\| w_k \right\|^{2} \right] & \leq  
			{\rho}^{2} \label{eqn: position noise 4} 
		\end{align}
		where $\rho $ is a positive constant. 
	\end{assumption} 
	
	\begin{remark}
		Assumption~\ref{asm: sensor noise} implies that the noises have zero mean and bounded variance. As a result, the noises are unbiased.  The noise variance is assumed to be bounded as in (\ref{eqn: sensor noise 2}) and \eqref{eqn: position noise 4}.	There are a number of reasons for the use of noisy variables.  First, in real experiment, the position where measurement is collected might not be exactly the same position as $ \bm p_k $. Possible causes are the error in the positioning system such as GPS for outdoor operation or SLAM (Simultaneous Localisation and Mapping) for indoor operation, or control errors in the lower loop. The autonomous search scheme acts as a higher level path planning and it passes the path or way points to a lower level controller to follow which may have control error due to disturbance and other reasons (e.g. local turbulence for a UAV operation).  Second, the concentration measurements collected are strongly subject to sensor noises and uncertainties. In this paper, we deploy additive white noise term to quantify those impacts, and we will analyse how the agent will behave under uncertain information. 
	\end{remark}

	In summary, the implementation structure of CL-DCEE can been encapsulated in Algorithm~\ref{alg: 1}. The key feature of our algorithm is the adoption of dual effects for exploitation and exploration. Multiple estimators have been developed for source term construction, which provides an effective tool for quantifying information uncertainty. In order to reduce future variance of the measurement, we need to predict future concentration by using current parameters of ATDM, based on which the exploration objective ${\mathcal P}_{k+1| k} $ can be formulated as a function of optimisation variable $ \bm p _{k+1|k}$. 
	\begin{algorithm}
		\textbf{Initialisation:}
		\begin{enumerate}
			\item[1.] specify the number of estimators $ N $
			\item[2.]  conduct a number of $q$ initial samples $(\bm p_{i},z(\bm p_{i}))$ indexed by $ i =-q+1, -q+2,  \dots, 0$
			\item[3.] initialise the agent's position $ \bm p_0 $ and the initial guess of the source parameter $ \bm\Theta_0^i $ for all $ i=1, 2, \dots, N $
		\end{enumerate}
		\textbf{Iteration:}
		\begin{enumerate}
			\item[4.] set $ k:=k+1 $
			\item[5.] collect the concentration reading from the sensor at position $\bm p_k $
			\item[6.] \textbf{for} $ i=1:N $\newline
			\hspace*{0.4cm}  update the estimated source terms by\\ 	
			\hspace*{0.4cm}	$\bm \Theta_{k}^i = \bm \Theta_{k-1}^i -  \sum_{t = k-q+1}^{k} \eta_t \tilde \nabla_{\bm \Theta}  f(\bm \Theta_{k-1}^i, \bm p_t)  $\\
			\textbf{end for}
			\item[7.] calculate current estimation variance \\ 
			\hspace*{0.4cm}	 $\mathcal P_{k| k} = \frac 1N \sum_{i=1}^N  ( z_k^i - \bar{ z}_{k} )^{2} $
			\item[8.] predict future variance as a function of $ \bm p_{k+1|k} $\\
			\hspace*{0.4cm} $  F_{k+1}^i = \left. \frac{\partial   \mathcal{M}\left(\bm {p}_{k+1|k},  {\bm\Theta}_k^i \right) }{\partial \bm p} \right |_{\bm p_{k+1|k}, {\bm \Theta}_k^i} $\\
			\hspace*{0.4cm}	$\mathcal P_{k+1|k} ={ \mathcal P}_{k|k} \bm F_{k+1}^{\tp} \bm F_{k+1}$
			\item[9.] update the next movement for the agent by\\
			\hspace*{0.4cm}  $\bm u_k =  - \delta_k \left[  \nabla_{\bm p}  y \left(\bm {p}_{k}, \bm \Theta_k \right)   +  \nabla_{\bm p} \mathcal P_{k+1|k}   \right] $\\
			\hspace*{0.4cm}  $\bm p_{k+1} =  \bm p_k + \bm u_k + w_k $
		\end{enumerate}
		\textbf{End if} termination condition is satisfied or iteration budget is approached. 
		\caption{Implementation structure of CL-DCEE}\label{alg: 1}
	\end{algorithm}

	\begin{remark}
		There is an important difference between the traditional gradient-based search methods such as in Chemotaxis and the proposed CL-DCEE algorithm in this study.  In the early works (e.g. \cite{Dhariwal2004bacterium, russell1995robotic}), mobile robots are equipped with sensors that directly collect the local gradients of concentration, and utilise the \emph{measured} gradients to plan their next movement. Clearly, this type of search suffers severely from sensor errors and turbulent fluctuations, since the next movement is purely determined by  instantaneous gradient measurements. In our framework, the search agent measures local concentration value, and uses all the available information, including priors and available measurements, to learn the source parameters. Based on the acquired knowledge of source, the search agent uses \emph{model evaluated} gradients to plan its next movement. This learning process lasts over the entire period of search, and therefore an instant sample, subject to noise and turbulence, will not cause considerable interruption to the path planning.  Combining with the multi-estimator scheme, the robustness of CL-DCEE algorithm is significantly improved, and is able to achieve comparable search performance as information-theoretic methods with less computational load, which will be demonstrated later through theoretical and simulation studies. 
	\end{remark}

	\subsection{Convergence Analysis}
	In this subsection, we will show that the path planning algorithm (\ref{alg: agent movement with dual obective}) in conjunction with multiple source estimators (\ref{alg: N source term gradient update}) will lead the agent to a small neighbourhood of the source location $ {\bm s} $.

	\begin{theorem}\label{thm: 1}
	Under Assumption~\ref{asm: sensor noise}, all $N$ source estimators in (\ref{alg: N source term gradient update}) converge to a neighbourhood of the true position of the release ${\bm s}$ from a random initialisation set if the learning rate $ \eta_t $ of each estimator is chosen such that 
		\begin{equation}\label{eqn: learning rate condition}
	\Gamma_k^i = \bigg \| I_4 -   \sum_{t = k-q+1}^{k} \eta_t  \bm{\mathcal T}_k^i(t)  \bigg\|^2 
    \end{equation}
    satisfies $0< \Gamma_k^i <1$, where
    $	\bm {\mathcal T}_k^i(t) :=  \int_{0}^{1} \nabla_{\bm \Theta}^{2} f ({\bm \Theta}_s +\tau \tilde{\bm \Theta}_k^i, \bm p_t  )d \tau. $
	Moreover, the expected mean-square-errors (MSE), $ \bE \|\bm \Theta^i_k - {\bm \Theta}_s\|^2 , \forall i =1, \dots, N$, converge at a geometric rate to a bounded neighbourhood of zero, given by 
	\begin{equation}\label{eqn: thm 1 MSE bound}
		\lim_{k\rightarrow \infty}  \bE \|\bm \Theta^i_k - {\bm \Theta}_s\|^2 \leq    \frac{\sup_{j \in[1, \infty)} ( \sum_{t = j-q+1}^{j} \eta_t^2 {\varrho}^{2} )}{1-\sup_{j \in[1, \infty)} (\Gamma_j^i )}.
	\end{equation} 
	\end{theorem}

	\begin{proof}
		It follows from (\ref{alg: N source term gradient update}) and (\ref{eqn: approximated gradient}) that 
		\begin{equation}\label{eqn: 12}
		\bm \Theta_{k+1}^i  = \bm \Theta_{k}^i - \sum_{t = k-q+1}^{k} \eta_t \left[ \nabla_{\bm \Theta} f(\bm \Theta_k^i, \bm p_t) + \mu_t  \right]
		\end{equation}
		Now, let $ \tilde{\bm \Theta}_k^i = \bm \Theta_k^i -  {\bm \Theta}_s $ denote the error of the agent's estimation relative to source parameters. Then, substituting $\tilde{\bm \Theta}_k$ into (\ref{eqn: 12}) results in the error dynamics as 
		\begin{equation}\label{eqn: 13}
		\tilde{\bm \Theta}_{k+1}^i = \tilde{\bm \Theta}_k^i - \sum_{t = k-q+1}^{k} \eta_t \left[\nabla_{\bm \Theta} f \left( {\bm\Theta}_k^i, \bm p_t \right) + \mu_t \right] .
		\end{equation}
		To relate the gradient term with $  \tilde{\bm \Theta}_k^i $, we resort to the mean value theorem~\cite{rudin1976principles}. For a twice-differentiable function $ H(x): \bR^m \rightarrow \bR $, the following relation holds, for any $ a,b \in \bR^m $,
		\begin{equation}\label{eqn: mean value theorem}
		\begin{aligned}
		\nabla_{x} H(b)=& \nabla_{x} H(a) \\ &+\left[\int_{0}^{1} \nabla_{x}^{2} H[a+\tau (b-a) ]d \tau \right](b-a).
		\end{aligned}
		\end{equation}
		Therefore, applying the above theorem leads to 
		\begin{equation}\label{eqn: 15}
		\begin{aligned}
		\nabla_{\bm \Theta} f\left(\bm {\Theta}_k^i, \bm p_t\right) =  & \nabla_{\bm \Theta} f\left({\bm \Theta}_s, \bm p_t \right) \\ & + \left[\int_{0}^{1} \nabla_{\bm \Theta}^{2} f({\bm \Theta}_s+\tau \tilde{\bm \Theta}_k^i, \bm p_t    ) d \tau\right] \tilde{\bm \Theta}_k^i 
		\end{aligned}
		\end{equation}
		Let us denote
		\begin{equation}\label{eqn: 17}
		\bm {\mathcal T}_k^i(t) :=  \int_{0}^{1} \nabla_{\bm \Theta}^{2} f ({\bm \Theta}_s +\tau \tilde{\bm \Theta}_k^i, \bm p_t  )d \tau. 
		\end{equation}
		Consequently, substituting (\ref{eqn: 17}) and (\ref{eqn: 15}) into (\ref{eqn: 13}) yields 
		\begin{equation}\label{eqn: 18}
		\tilde{\bm \Theta}_{k+1}^i = \bigg(I_4 -  \sum_{t = k-q+1}^{k}  \eta_t \bm{\mathcal T}_k^i(t) \bigg)	\tilde{\bm \Theta}_{k}^i - \sum_{t = k-q+1}^{k} \eta_t\mu_t
		\end{equation}
		where $\nabla_{\bm \Theta} f\left({\bm \Theta}_s, \bm p_t \right) = \bm 0$ has been used. 
		Taking the square of the Euclidean norm  of the error dynamics (\ref{eqn: 18}) gives 
		\begin{equation}\label{eqn: 19}
		\begin{aligned}
		\| \tilde{\bm \Theta}_{k+1}^i \|^2  = & \bigg\|  \bigg(I_4 -  \sum_{t = k-q+1}^{k} \eta_t   \bm{\mathcal T}_k^i(t) \bigg )\tilde{\bm \Theta}_k^i -  \sum_{t = k-q+1}^{k} \eta_t  \mu_t \bigg \|^2 \\
		= & \bigg \| \bigg(I_4 -  \sum_{t = k-q+1}^{k}\eta_t   \bm{\mathcal T}_k^i(t) \bigg)\tilde{\bm \Theta}_k^i  \bigg \|^2 \\ & + \sum_{t = k-q+1}^{k} \eta_t^2 \left \| \mu_t \right \|^2  \\ & -2   \bigg[(I_4 -  \sum_{t = k-q+1}^{k} \eta_t  \bm{\mathcal T}_k^i(t) )\tilde{\bm \Theta}_k^i\bigg]^{\tp}   \sum_{t = k-q+1}^{k} \eta_t \mu_t .
		\end{aligned}
		\end{equation}
	
%
		Let $ \mathbb{Q}_k^i := \bE\| \tilde{\bm{\Theta}}_k^i \|^2 $ denote the expected mean-square-error of the  variable $ \tilde{\bm{\Theta}}_k  $. Then, taking the expectation of (\ref{eqn: 19}) results in 
		\begin{equation}\label{eqn: P inequality 1}
		\begin{aligned}
		\mathbb{Q}_{k+1}^i \leq & \bigg \| I_4 -   \sum_{t = k-q+1}^{k} \eta_t  \bm{\mathcal T}_k^i(t) \bigg\|^2  \mathbb{Q}_{k}^i +\sum_{t = k-q+1}^{k} \eta_t^2 {\varrho}^{2} \\
            &-2 \bE  \bigg[(I_4 -  \sum_{t = k-q+1}^{k} \eta_t  \bm{\mathcal T}_k^i(t) )\tilde{\bm \Theta}_k^i\bigg]^{\tp}  \sum_{t = k-q+1}^{k} \eta_t \mu_t   \\
            = & \bigg \| I_4 -   \sum_{t = k-q+1}^{k} \eta_t  \bm{\mathcal T}_k^i(t) \bigg\|^2  \mathbb{Q}_{k}^i +\sum_{t = k-q+1}^{k} \eta_t^2 {\varrho}^{2}
		\end{aligned}
		\end{equation} 
		where conditions of the gradient noise in Assumption~\ref{asm: sensor noise} have been utilised, i.e., $ \mu_t $ is white noise independent of $\bm \Theta_k^i$ with bounded variance.
		To guarantee the convergence of the estimators, it is required that 
		\begin{equation}\label{eqn: convergence iteration}
			\Gamma_k^i = \bigg \| I_4 -   \sum_{t = k-q+1}^{k} \eta_t  \bm{\mathcal T}_k^i(t)  \bigg\|^2 
		\end{equation}
	    within unit circle. 
		Then, we have 
		\begin{equation}\begin{aligned}
		\lim_{k\rightarrow \infty} \mathbb{Q}_k^i &    \leq \frac{\sup_{j \in[1, \infty)} ( \sum_{t = j-q+1}^{j} \eta_t^2 {\varrho}^{2} )}{1- \sup_{j \in[1, \infty)}  (\Gamma_j^i )}
		\end{aligned}
		\end{equation}
		where $\lim_{k\rightarrow \infty} (\prod_{j=1}^{k} \Gamma_j^i ) \mathbb{Q}_{0}^i= 0 $ has been applied. 
		In view of (\ref{eqn: P inequality 1}), it can be concluded that the estimator MSE converges to a small neighbourhood of zero at a geometric rate, given by $ \mathcal O(\sup_{j \in[1, \infty)} \Gamma_j^i )$.
		This completes the proof.
	\end{proof}

		\begin{remark}\label{rem: 5}
			To ensure that $\Gamma_k^i $ in \eqref{eqn: learning rate condition} is within unit circle, it is sufficient to require $\sum_{t = k-q+1}^{k} \bm{\mathcal T}_k^i(t)>0 $ for a positive integer $q$ under small learning rate $\eta_t>0$.  This is a commonly-used condition of persistent excitation in adaptive control and system identification \cite{ortega2020modified,guay2003adaptive}. In essence, at each iteration $k$, not only current data sample $(\bm p_k, z(\bm p_k)) $ is utilised for updating the environment parameter $\bm \Theta_{k+1}$ but also past $q-1$ step measurements $(\bm p_j, z(\bm p_j)) $ for $j\in\{k-q+1,\dots, k-1\}$ are used. This type of technique is motivated by the memory regressor extension, see \cite{ortega2020modified}, to relax the requirement of persistent excitation for each single sample. 
		\end{remark}
	\begin{remark}\label{rem: 6}
		The proposed parameter adaption algorithm in \eqref{alg: N source term gradient update} encompasses two special cases commonly-used in existing literature: stochastic gradient approximation ($q=1$) and full batch approximation ($q=k$). It is worth noting that increasing the iteration length $q$ can enhance the robustness and accuracy of the adaption algorithm as the excitation effect will be more significant, but may also incur additional computational load \cite{ding2007performance}.  
	\end{remark}

\begin{remark}\label{rem: 7}
	Different from existing filtering techniques, like extended Kalman filter and Gaussian mixture filter, which usually rely on process models and stochastic properties of process noises to quantify the level of estimation uncertainty, the proposed concurrent learning method in this paper uses a hybrid approach that combines both model-based and model-free techniques. The model-based parallel estimators essentially yield a distribution of the estimation at each iteration. A model-free approach is used to calculate the mean of the estimation and its variance based on the distribution of the estimations yielded by these parallel estimators. Recently, this hybrid model-based and model-free approach has been proven to be very successful and promising via extensive simulation and experimental studies \cite{chua2018deep, Lakshminarayanan2017simple} in machine learning community. It takes the advantage of the model-based approaches in sampling efficiency but alleviates its inherited model biased error using a model-free ensemble. However, there is no rigorous result for the ensemble approach in machine learning community despite its widely perceived success. Inspired by its success in machine learning, we propose a hybrid parameter estimation approach consisting of $N$ parallel gradient based estimators and an ensemble process. This approach not only significantly increases the robustness of the parameter estimation particularly in the presence of intermittent sensor measurement but also provides a reliable way to quantify the level of uncertainty of the current estimation, which is important in realising the DCEE concept. In Theorem~1, we prove its global convergence of the estimation for our specific application by leveraging a memory based estimation method.
	\end{remark}

	Theorem~\ref{thm: 1} shows the convergence of the estimators, i.e., the estimator will eventually converge using feasible path planning methods, but the optimality is not guaranteed. Convergence of source estimation can be achieved as long as the agent keeps collecting information that fulfils the conditions specified in Theorem~\ref{thm: 1}. In a real search problem, the search environment is complex and there is limited time/sampling budget, and therefore the search agent has to actively plan its path to quickly approach the source.
	
	Although the path planning and environment acquisition are coupled, it has been shown that under Assumption~\ref{asm: sensor noise} source estimators can converge to true parameters when measurement samples $ k\rightarrow \infty $. This important property allows us to employ the well-known separation principle for the convergence analysis of the overall algorithm. Such an analytical principle has been widely used to establish the stability of disturbance observer based control (DOBC) \cite{chen2004disturbance, li2011generalized}, where design of the controller is separated from design of the observer. In addition, we will further analyse the composite search performance (steady-state performance) in relation to the noise characters.

	\begin{theorem}\label{theorem}
		Consider a dispersion described by ATDM (\ref{eqn: ATD model}) and the measurement errors and disturbances satisfy Assumption~\ref{asm: sensor noise}. Let $\eta_t$ satisfy the condition specified in Theorem \ref{thm: 1}.  If the step size $ \delta_k $ is designed such that
		\begin{equation}\begin{aligned}
             0< 2 \| I_3 - \delta_k \bm{\mathcal L}_k \|^2  <1
			\end{aligned}
		\end{equation}
	where 	$		\bm {\mathcal L}_k :=  \int_{0}^{1} \nabla_{\bm p}^{2} y ({\bm s }+\tau \tilde{\bm p}_k, \bm \Theta_k  )d \tau $,
	then the search agent converges to a bounded neighbourhood of the source location using the proposed CL-DCEE in Algorithm~\ref{alg: 1}. Moreover, the steady-state MSE bound between agent and true source is given by 
		\begin{equation}\begin{aligned}
		\lim_{k\rightarrow \infty} & \bE \|\bm p_k - \bm s \|^2
		\leq  
		\frac{ \bar \nu^2  + {\varrho}^{2} }{1- \sup_{j \in[1, \infty)}(2 \| I_3 - \delta_j \bm{\mathcal L}_j \|^2 )}
	\end{aligned}
    \end{equation}
	where $ \bar \nu >0$ denotes the upper bound of the gradient norm of the estimators' variance $\| \nabla_{\bm p} \mathcal P_{k+1|k} \|$.
	\end{theorem}
	\begin{proof}
    According to the path update law (\ref{alg: agent movement with dual obective}), we have  
		\begin{equation}\label{eqn: 43}
		\bm p_{k+1} = \bm p_k - \delta_k \left[\nabla_{\bm p} y \left(\bm {p}_{k},\bm \Theta_k \right) +  \nabla_{\bm p} \mathcal P_{k+1|k}  \right] +w_k
		\end{equation}
		Denote $ \tilde{\bm p}_k = \bm p_k -  {\bm s} $ as the error of the agent's position relative to the source position. Consequently, the error dynamics of $ \tilde{\bm p}_k$  can be written as
		\begin{equation}\label{eqn: 44}
		\tilde{\bm p}_{k+1} = \tilde{\bm p}_k -\delta_k \nabla_{\bm p} y\left(\bm {p}_{k},\bm \Theta_k \right)- \delta_k \nabla_{\bm p} \mathcal P_{k+1|k} +w_k.
		\end{equation}
		Following a similar argument as in Theorem~\ref{thm: 1}, we have 
		\begin{equation}\label{eqn: 45}
		\tilde{\bm p}_{k+1} = (I_3 - \delta_k \bm{\mathcal L}_k)\tilde{\bm p}_k- \delta_k \nabla_{\bm p} \mathcal P_{k+1|k} +w_k
		\end{equation}
		where 
		\begin{equation}\label{eqn: 46}
		\bm {\mathcal L}_k :=  \int_{0}^{1} \nabla_{\bm p}^{2} y ({\bm s }+\tau \tilde{\bm p}_k, \bm \Theta_k  )d \tau.
		\end{equation}
		Then, taking the square of the Euclidean norm for both sides of the error dynamics (\ref{eqn: 45}) leads to 
		\begin{equation}\label{eqn: 47}
		\begin{aligned}
		\| \tilde{\bm p}_{k+1} \|^2  = & \| (I_3 - \delta_k \bm{\mathcal L}_k)\tilde{\bm p}_k - \delta_k \nabla_{\bm p} \mathcal P_{k+1|k} +w_k \|^2 \\
		= & \| (I_3 - \delta_k \bm{\mathcal L}_k)\tilde{\bm p}_k \|^2  +\delta_k^2 \| \nabla_{\bm p} \mathcal P_{k+1|k} \|^2  +  \| w_k \|^2 \\ & + 2  [ (I_3 - \delta_k \bm{\mathcal L}_k)\tilde{\bm p}_k ]^{\tp}w_k  \\ & - 2 \delta_k   \nabla_{\bm p}^{\tp} \mathcal P_{k+1|k}w_k   \\ & -2 \delta_k [ (I_3 - \delta_k \bm{\mathcal L}_k)\tilde{\bm p}_k ]^{\tp}  \nabla_{\bm p} \mathcal P_{k+1|k} .
		\end{aligned}
		\end{equation}
		Let $ \mathbb{P}_k := \bE\| \tilde{\bm{p}}_k \|^2 $ denote the expected mean-square-error between the agent's position and the source location. Taking the expectation of (\ref{eqn: 47}) and further applying the noise conditions in \eqref{eqn: position noise 3} and \eqref{eqn: position noise 4}, we have 
		\begin{equation}\label{eqn: P inequality 48}
		\begin{aligned}
		\mathbb{P}_{k+1} \leq &  \| I_3 - \delta_k \bm{\mathcal L}_k \|^2  \mathbb{P}_{k} +  \bE[ \delta_k^2 \| \nabla_{\bm p} \mathcal P_{k+1|k} \|^2 ] + {\rho}^{2}  \\
	 	& + \bE[ 2  [ (I_3 - \delta_k \bm{\mathcal L}_k)\tilde{\bm p}_k ]^{\tp}w_k ] \\ & - \bE[2 \delta_k  \nabla_{\bm p}^{\tp} \mathcal P_{k+1|k}w_k  ]\\
		& - \bE[ 2 \delta_k  [(I_3 - \delta_k \bm{\mathcal L}_k)\tilde{\bm p}_k ]^{\tp} \nabla_{\bm p} \mathcal P_{k+1|k}  ] \\
		\leq &	2 \| I_3 - \delta_k \bm{\mathcal L}_k \|^2  \mathbb{P}_{k} +  {\rho}^{2} \\ & + 2   \delta_k^2 \| \nabla_{\bm p} \mathcal P_{k+1|k} \|^2 
		\end{aligned}
		\end{equation} 
		where the following three relationships have been applied to derive the second inequality,
		\begin{equation}
		\begin{aligned}
		& \bE[ 2  [ (I_3 - \delta_k \bm{\mathcal L}_k)\tilde{\bm p}_k ]^{\tp}w_k ] =0 \\ & \bE[2 \delta_k  \nabla_{\bm p}^{\tp} \mathcal P_{k+1|k}w_k  ]  = 0 \\
		& \bE [ 2 \delta_k [(I_3 - \delta_k \bm{\mathcal L}_k)\tilde{\bm p}_k ]^{\tp}  \nabla_{\bm p} \mathcal P_{k+1|k}  ] \\ & \ \ \ \leq \| \delta_k \nabla_{\bm p} \mathcal P_{k+1|k} \|^2+  \bE \| (I_3 - \delta_k \bm{\mathcal L}_k)\tilde{\bm p}_k \|^2 \\ & \ \ \ =  \delta_k^2 \| \nabla_{\bm p} \mathcal P_{k+1|k} \|^2  + \| I_3 - \delta_k \bm{\mathcal L}_k \|^2  \mathbb{P}_{k} .
		\end{aligned}
		\end{equation}
		In view of the definition of $\mathcal P_{k+1|k}$, it is known that the last term in (\ref{eqn: P inequality 48}), $ 2  \delta_k^2 \| \nabla_{\bm p} \mathcal P_{k+1|k} \|^2  $, is a measure of the variance of the estimator error that is upper bounded by 
		\begin{equation}\label{upper bound of estimation error}\begin{aligned}
			\bE \| \tilde{\bm \Theta}_{k}^i \|^2  \leq 	\max \left\{\| \tilde{\bm \Theta}_{0}^i \|^2 , \frac{\sup_{j \in[1, \infty)}( \sum_{t = j-q+1}^{j} \eta_t^2 {\varrho}^{2} )}{1-\sup_{j \in[1, \infty)}(\Gamma_j^i )} \right\}
			\end{aligned}
		\end{equation}
	where $\| \tilde{\bm \Theta}_{0} \|^2 $ is the initial estimation error of the estimators.
		Note that the ATDM is a smooth function with respect to the source estimators ${\bm \Theta}_k^i$, and thus $ \bm F_{k+1} $ is bounded for bounded ${\bm \Theta}_k^i$, as in \eqref{upper bound of estimation error}. Therefore, we can always find an upper bound $ \bar \nu^2 >0 $ such that $   \|\nabla_{\bm p} \mathcal P_{k+1|k} \|^2\leq\bar \nu^2 $.
    Thus, 
		\begin{equation}\label{eqn: 50}
		\mathbb{P}_{k+1} \leq  (2 \| I_3 - \delta_k \bm{\mathcal L}_k \|^2  )\mathbb{P}_{k} + \bar \nu^2  + {\varrho}^{2}   .
		\end{equation}
	If we choose $\delta_k$ such that $(2 \| I_3 - \delta_k \bm{\mathcal L}_k \|^2  ) $ is within unit circle, then the convergence of \eqref{eqn: 50} is guaranteed.

		Now, we analyse the steady-state search performance. 
		It follows from (\ref{eqn: 50})  that 
		\begin{equation}\label{eqn: position MSE}\begin{aligned}
		\lim_{k\rightarrow \infty} & \bE \|\bm p_k - \bm s \|^2
		\leq  
        \frac{ \bar \nu^2  + {\varrho}^{2} }{1- \sup_{j \in[1, \infty)}(2 \| I_3 - \delta_j \bm{\mathcal L}_j \|^2 )}
		\end{aligned}
		\end{equation}
		where $\lim_{k\rightarrow\infty}\prod_{j=1}^{k}( 2 \| I_3 - \delta_j \bm{\mathcal L}_j \|^2  ) \mathbb{P}_{0} = 0 $ has been applied. 
		Similarly, it can be obtained from (\ref{eqn: 50}) that the agent converges to a bounded mean-square-error in (\ref{eqn: position MSE}) at a geometric rate, given by $ \mathcal O(\sup_{j \in[1, \infty)}(2 \| I_3 - \delta_j \bm{\mathcal L}_j \|^2 )) $.
		This completes the performance analysis.
	\end{proof}

	\begin{remark}
	There is a significant difference between the existing dual control formulation and our framework in this paper. Previous studies introduce the exploration effect on the \emph{system} for purposes of state or parameter estimation \cite{mesbah2018stochastic, bugeja2009dual, filatov2000survey}, while in our work the probing effect is used to explore the \emph{environment} (in this case, learn the source location and release rate). This crucial distinction allows us to learn the unknown environment by reducing estimation uncertainty.  Compared with our previous work in \cite{Chen2021DCEE}, there are several distinctions in this paper. a) The problem formulation is different. The formulation in (10) is a concentration-driven optimisation problem, whereas \cite{Chen2021DCEE} uses a position-driven mechanism (see equation (5) in \cite{Chen2021DCEE}). This subtle distinction leads to different search behaviours and analytical procedures. From the traditional search strategies point of view, one relates to Chemotaxis while the previous one to Infotaxis \cite{Hutchinson2017review}. b) DCEE \cite{Chen2021DCEE} uses particle filters for the source term estimation and also for posterior estimation, which is quite computationally expensive. We moved away form this framework to reduce computational burden to make autonomous algorithms easily implemented on mobile sensor platforms that normally have limited computational resources. Motivated by ensemble aggregating in machine learning community \cite{chua2018deep, Lakshminarayanan2017simple}, a concurrent learning mechanism for the estimation process is proposed using a multi-estimator ensemble approach.  It is shown that this learning based control method is much more computationally efficient. c) The feasible action set $\Omega$ in this paper can be continuous, whereas in \cite{Chen2021DCEE} only a limited number of feasible actions can be chosen. Limited discretised actions in \cite{Chen2021DCEE} may heavily influence the flexibility of the search agent and restrict the potential of dual control. d) In this work, we provide a complete theoretical analysis of the modified dual control algorithm using gradient descent. There is no theoretic analysis of convergence property of DCEE in \cite{Chen2021DCEE}. In fact, so far there is no formal analysis and theoretic proof for any IPP based search algorithms including DCEE, which is even more complicated due to the involvement of dual affect.
	\end{remark}

	\begin{remark}
		If we remove the second term $\mathcal P_{k+1| k}$ in the path planning objective in (\ref{eqn: N dual objective}), then our algorithm reduces to the pure exploitation strategy, which solely relies on the current estimators of the source parameters. It should be emphasised that the learning process of pure exploitation is \emph{passive} or accidental, since source parameters are updated when the agent makes full use of current belief. In Algorithm~\ref{alg: 1}, the probing effect is included in the value function, by which the agent can \emph{actively} or deliberately learn the environment. In this sense, our CL-DCEE framework is closely related to active learning in MPC \cite{barshalom1974dual, mesbah2018stochastic, Chen2021DCEE}. Generally speaking, dual control of exploration and exploitation in an uncertain environment belongs to a much wider class of machine learning problems, in particular, reinforcement learning \cite{fisac2018general, jeong2019learning, guha2020MRAC}. 
	\end{remark}

	\section{Simulation Study}\label{sec: 4}
	In this section, simulation results will be provided to validate the effectiveness of the proposed algorithms. Since Entrotaxis \cite{Hutchinson2018Entrotaxis} has demonstrated better performance compared with other existing methods, we will use Entrotaxis as a benchmark for the simulation study. It is worth noting that those informative path planing approaches require a significant amount of computational power due to the implementation of the nonlinear Bayesian filtering and the sampling search based path planning structure. 
	
	An autonomous UAV is utilised to search an open bounded space where a single source release is present. The area of interest is $\Omega = 100\text{m}\times 100\text{m} \times1 $m. For the first experiment in Section \ref{sec: 4.1}, environmental parameters are known: the wind speed $u_s= 4 \rm{m/s}$, wind direction $\rho_{s} =  1.5\pi $ rad, diffusivity $\zeta_{s1}= 1 $, the particle lifetime $\zeta_{s2}= 20$. For the second case in Section \ref{sec: 4.2}, the source and environmental parameters are all considered to be unknown. The operational parameters of the UAV are listed in Table \ref{tab: 1}. In order to achieve stable sensor measurement, the UAV will take $10$s to collect one concentration sample. The maximum sampling and flight time budgets are bounded. Searching process will be terminated when the maximum budgets are reached.  True source parameters and prior knowledge used by the agent are summarised in Table~\ref{tab: 1}, where $ U(\cdot) $ denotes uniform distribution, and $ N(\cdot) $ represents normal distribution.

	\begin{table}
		\centering
		\caption{Operational parameters and source knowledge.}
		\label{tab: 1}
		\begin{tabular}{@{}llll@{}}
			\toprule
			&  UAV  	&  Source & Prior  \\ \midrule
			Measurement budget & $180$ & - & - \\ 
			Flight budget& $2,000$s & -& - \\
			Velocity & $2$m/s  & -& - \\
			Maximum step size & $4$m & -&- \\
			Start position & $[2,2,1] $ & -& - \\
			$x$ position &  -& $80$m & $U(x_{min}, x_{max})$   \\ 
			$y$ position &  -& $80$m & $U(y_{min}, y_{max})$   \\ 
			$z$ position &  - &$1$m  & $U(z_{min}, z_{max})$   \\ 
			Release rate  &  - & $10$g/s & $N(11,2)$ 
			\\ \bottomrule
		\end{tabular}
	\end{table}

	The algorithm structure of Entrotaxis is presented in  \cite{Hutchinson2018Entrotaxis}, which is composed of an inference engine and a path planner. The number of particles for the inference engine is set as $10,000$, and feasible directions are  $\mathcal{U} =\{\uparrow, \downarrow, \leftarrow, \rightarrow, \nwarrow, \nearrow, \swarrow, \searrow\} $ with a fixed step size $ 2 $m. For the proposed method in this paper, we implement the CL-DCEE framework in Algorithm~\ref{alg: 1} with six different sets of environment estimators, $ N=5, 10, 50, 100, 200$ and $1,000$, respectively. Different from Entrotaxis and DCEE, the step size of movement can be any value in the range of $ [1\rm{m},4\rm{m}] $, and the direction of movement can be arbitrary by taking advantage of low computational load of CL-DCEE. The learning rates are set as $ \eta_k =5 $ and $\delta_k =4$. Since one step stochastic approximation generates satisfactory estimation performance with high efficiency, the memory integer $ q$ is set as 1.  
	The initial conditions of estimators are randomly chosen in accordance with the prior knowledge in Table \ref{tab: 1}. 
	It should be noted that measurement signals from sensors are converted in dB scale.
	When implementing the proposed algorithms, the gradients are normalised to keep agent's movement within a meaningful range.

	\subsection{Unbiased Sensor Noises in Partially Unknown Environment}\label{sec: 4.1}
	In this subsection, the sensor measurement noises are set as additive Gaussian white noises. 
	We assume that there is no non-detection event, which means the sensor can always obtain concentration values buried in noises. Each algorithm has been repeated for $200$ times with the same configurations. 
	
\begin{figure}
		\centering
		\includegraphics[width=1\linewidth]{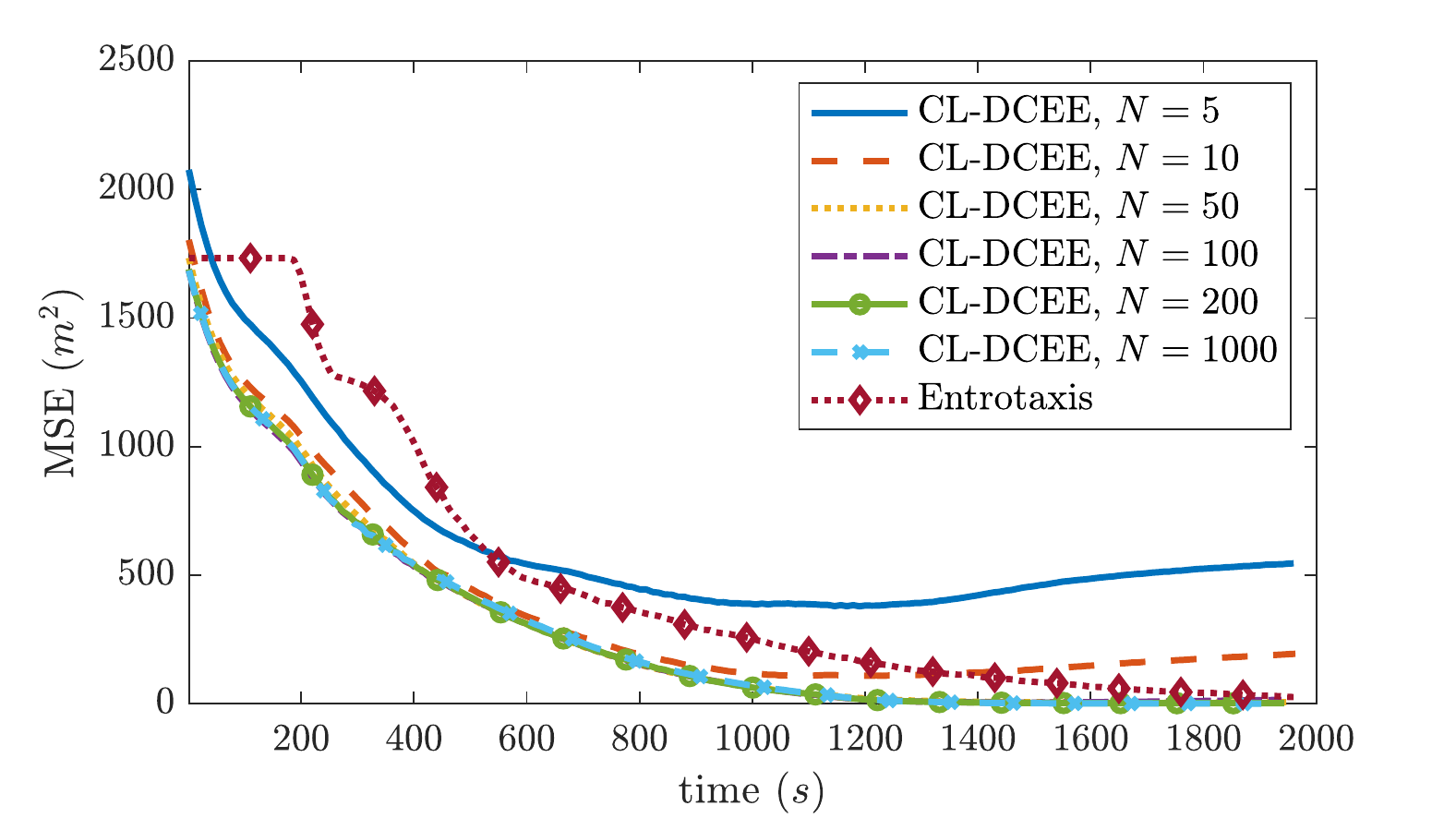}
		\caption{Mean-square-error between the estimated and true source positions.}
		\label{fig: 1}
	\end{figure}

	The obtained mean-square-errors of the CL-DCEE and Entrotaxis algorithms are displayed in Fig.~\ref{fig: 1}. MSE evaluates the performance of the source estimators, calculated by $  \bE(\bar{\bm s}_{k} -\bm s )^2 $. It is clear that all  algorithms can gradually achieve acceptable estimation of the source position within limited budgets. Uniform distribution of the source location has been applied in the initialisation process, as it is assumed that there is no prior information regarding source position. As a result, the initial guess of the source position is around the centre of the search space. In general, Entrotaxis requires a large number of measurements to update its particle filter, which leads to a slow acquisition rate of source estimation. The acquisition process continues until the flight is terminated at the maximum measurement budget. On the other hand, our proposed algorithms allow quick update of the source estimators by using instantaneous measurements. CL-DCEE algorithm converges to bounded MSE at approximately $1,000$s. This property is helpful to conducting emergent identification of the source parameters.

	\begin{figure}
		\centering
		\includegraphics[width=1\linewidth]{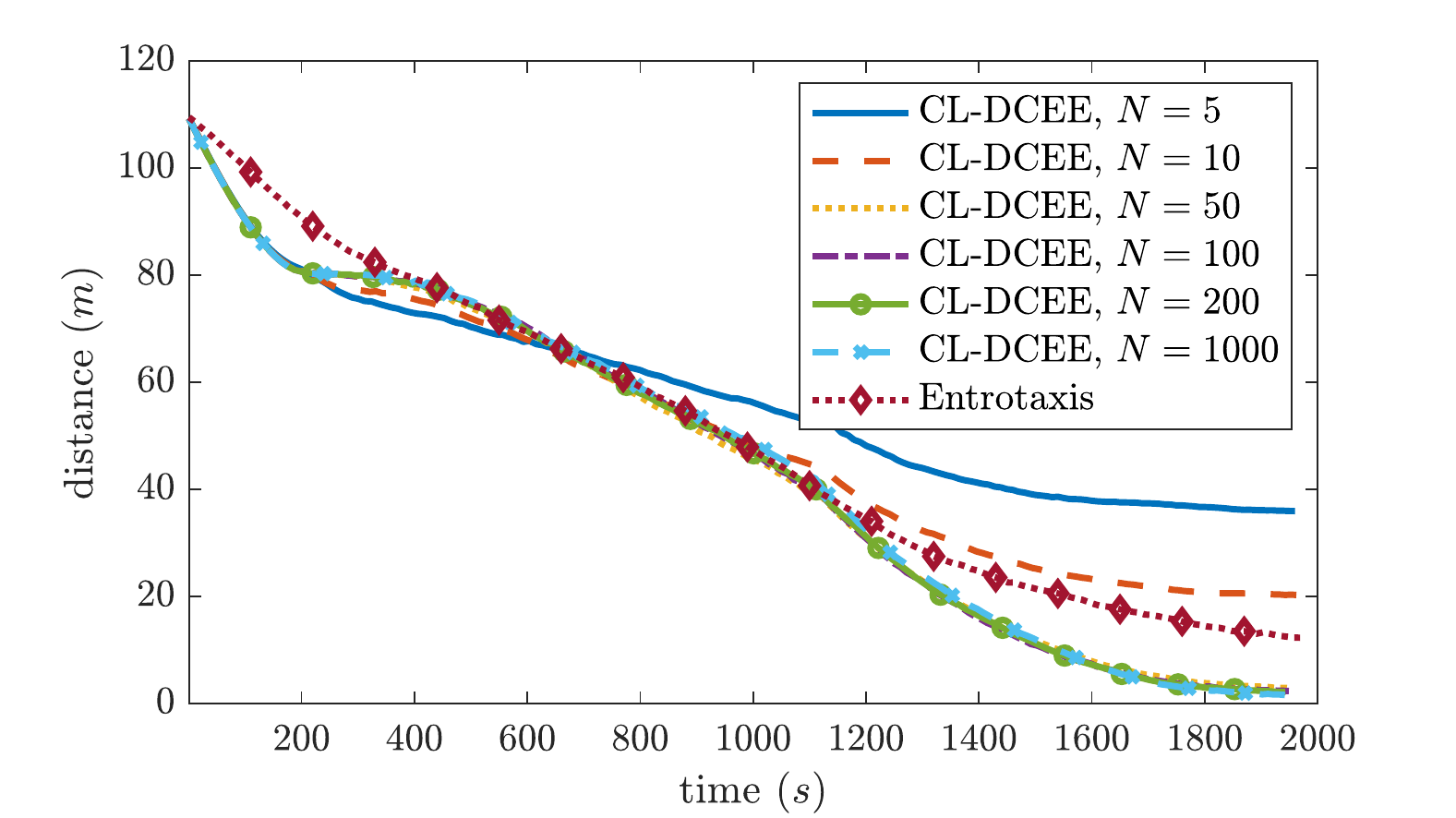}
		\caption{Distance between agent's position and the true source.}
		\label{fig: 2}
	\end{figure}

	\begin{figure*}
		\centering
		\begin{subfigure}{1\textwidth}
			\centering
			\includegraphics[width=1.0\linewidth]{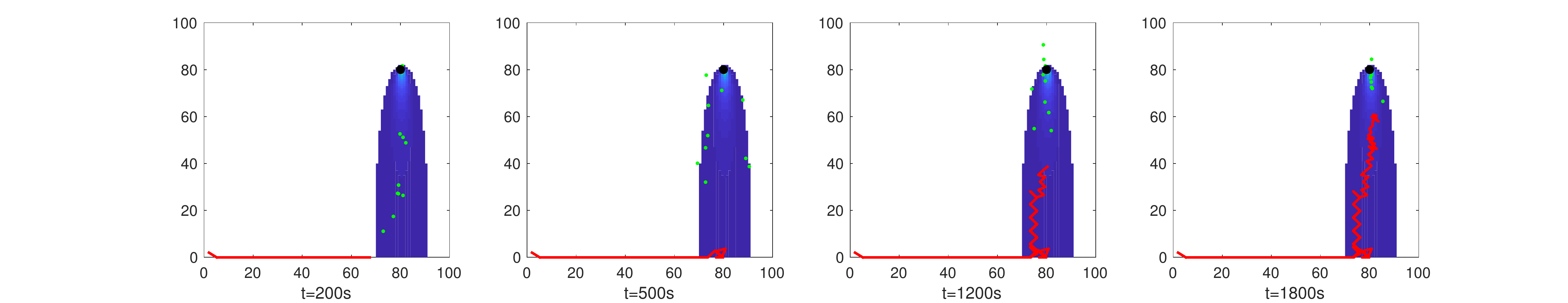}
			\caption{}
			\label{fig: 3a}
		\end{subfigure}
		
		\begin{subfigure}{1\textwidth}
			\centering
			\includegraphics[width=1.0\linewidth]{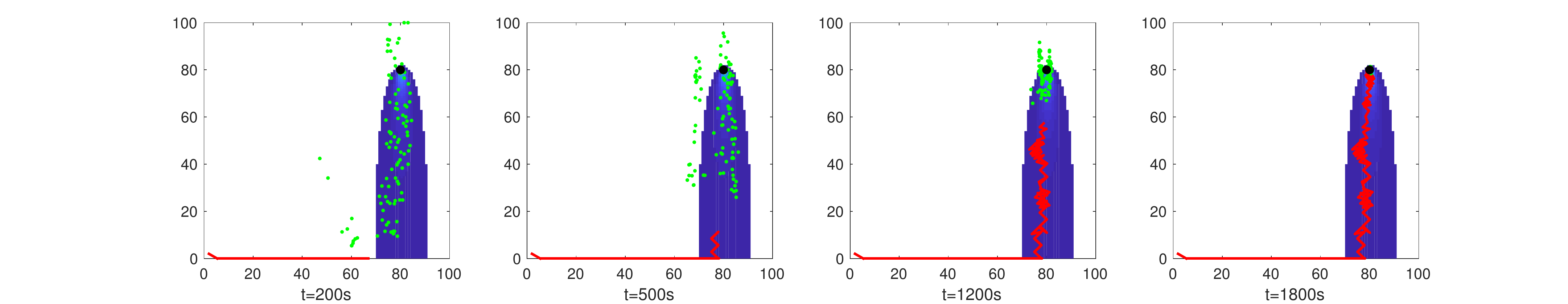}
			\caption{}
			\label{fig: 3b}
		\end{subfigure}
		
		\begin{subfigure}{1\textwidth}
			\centering
			\includegraphics[width=1.0\linewidth]{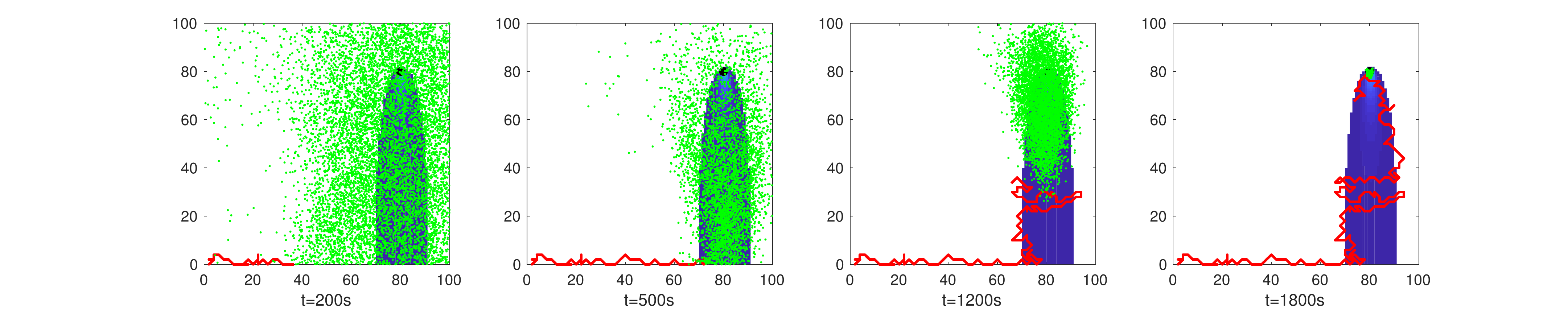}
			\caption{}
			\label{fig: 3c}
		\end{subfigure}
		\caption{Representative runs of different algorithms: (a) CL-DCEE with $N=10$, (b) CL-DCEE with $N=100$, (c) Entrotaxis. Red lines are the paths of the UAV, the green dots represent the estimated source position, and the black dots represent the true source position. }
	\end{figure*}
	
	\begin{figure}
		\centering
		\includegraphics[width=1\linewidth]{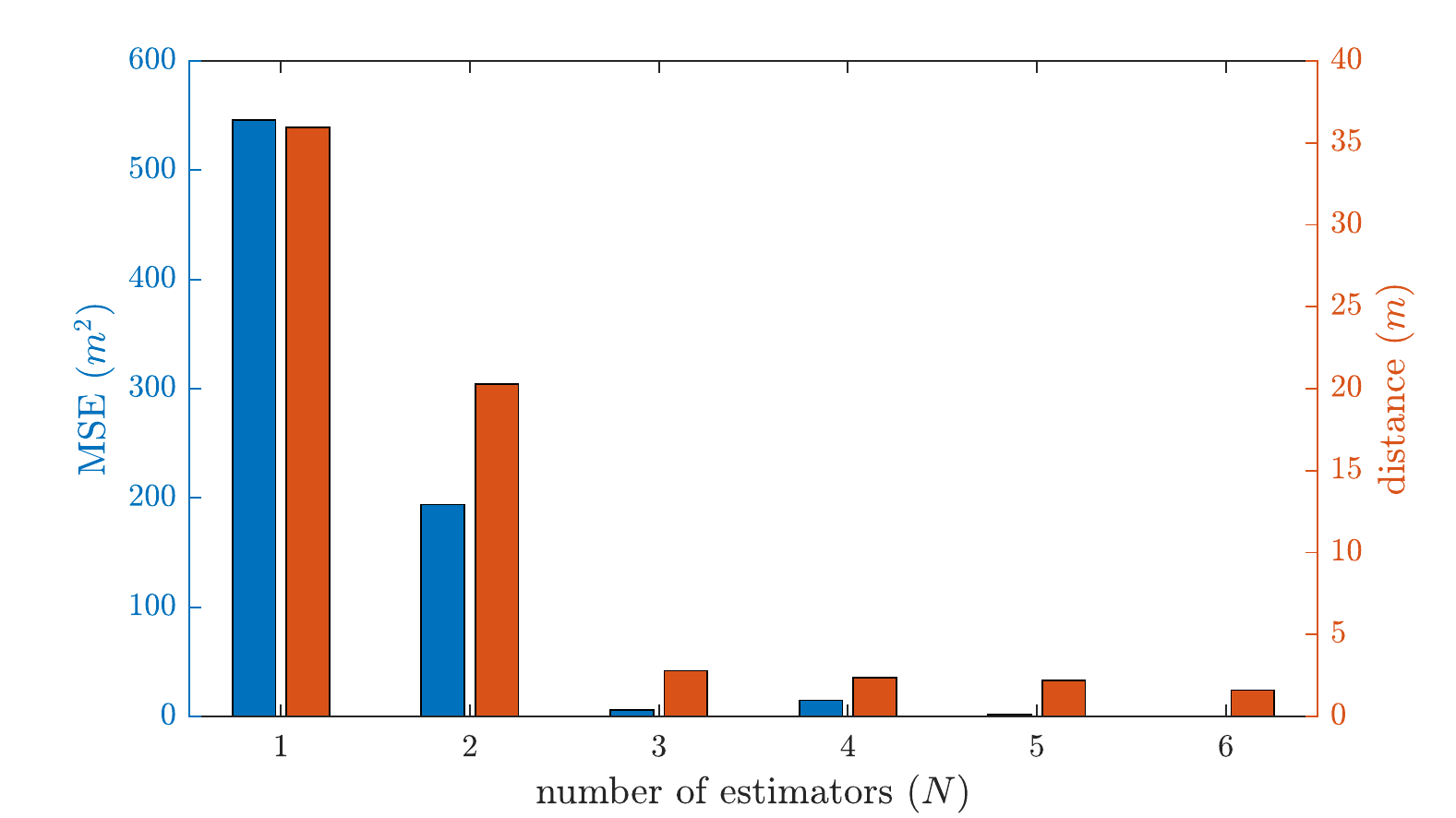}
		\caption{Performance of CL-DCEE algorithm with different number of estimators.}
		\label{fig: 3}
	\end{figure}
	
	Apart from the estimation accuracy, it is also desired that the search agent can move to the source position, so as to closely monitor the status of the release or take further remedy actions. In Fig. \ref{fig: 2}, the distance between the agent and the source is displayed. A noticeable phenomenon is that agent's position using Entrotaxis is quite far from the source position. The proposed CL-DCEE in this paper can keep the agent in the neighbourhood of the true source, and the steady-state distances are around $35.96$m, $20.28$m, $2.80$m, $2.36$m, $2.22$m and $1.61$m, for $N = 5, 10, 50, 100, 200$ and $1,000$, respectively.  Illustrative examples of search paths of CL-DCEE with $N=10$ and $N=100$ have been demonstrated in Figs. \ref{fig: 3a} and \ref{fig: 3b}.

	To show the influence of the number of estimators, we have presented the average performance using different values of $ N $, as shown in Fig. \ref{fig: 3}. Initially, increasing $ N $ can significantly enhance the performance in terms of estimators' MSE and the agent distance to the source ($N$ ranging from $5$ to $50$). 
	For $ N\geq50 $, increasing $N$ is no longer able to provide much performance improvement ($N$ ranging from $50$ to $1,000$). Therefore, the proposed CL-DCEE framework does not require a large number of estimators, and tens of them will be sufficient for autonomous search problem. It also implies that the number of estimators for the ensemble approach should be properly selected to balance estimation performance and computational complexity.

	Entrotaxis aims to minimise the estimation error of source parameters, and it does not necessarily imply that the agent has to move close to the source position. Correspondingly, there is a substantial distance between the agent and the source, even though the estimated source position has been fairly satisfactory. A representative search path  using Entrotaxis has been presented in Fig.~\ref{fig: 3c}. In essence, Entrotaxis and other informative path planning methods utilise pure exploration strategy \cite{Chen2021DCEE}, which drives the agent to the most informative position in order to reduce information uncertainties. As can be seen from~Fig. \ref{fig: 3c}, the agent ignores the current estimation of the source position but probes the uncertain areas around the estimated source. This consequently leads to large errors between the agent and the source.

	An important advantage of the proposed methods in this paper is the computational efficiency. For clear comparison, we have summarised time consumed by different algorithms, as shown in Table \ref{tab: 2}. The simulations are carried out using Matlab with a processor of 2.8 GHz Quad-Core Intel Core i7. It can be seen that our algorithm is much faster than Entrotaxis. It only consumes less than $1\%$ of the time for Entrotaxis ($N\leq 100$).  As a result, CL-DCEE also occupies much less memory storage since the number of estimators is much smaller. This is a very important and advantageous feature because processors used on mobile platforms are usually lower-price portable chips that cannot offer intensive computational power or large memory.

	\begin{table}
		\setlength\heavyrulewidth{0.25ex}
		\centering
		\caption{Time consumed by running different algorithms for $200$ trials.}
		\label{tab: 2}
		\resizebox{\linewidth}{!}{%
			\begin{tabular}{@{}cccccccc@{}}
				\toprule
				& \multicolumn{6}{c}{CL-DCEE}      & Entrotaxis \\ \midrule
				Estimators/Particles & 5 & 10 & 50 & 100 & 200 & 1,000 & 10,000      \\
				Time (second)        &21.8   & 23.1   & 24.3   &26.3    &30.5     & 56.8      &   2940.4
				\\ \bottomrule
			\end{tabular}%
		}
	\end{table}

	\subsection{Unknown Environment with Sensor Non-detection Events}\label{sec: 4.2}
	In this subsection, operational environment is also assumed to be unknown. Consequently, this necessitates estimating a full list of parameters of ATDM, i.e. $\bm \Theta_s =\left[\bm{s}^{\tp}, q, \rho_{s}, u_{s}, \zeta_{s 1}, \zeta_{s 2}\right]^{\tp}$. 
	One of the most significant challenges in autonomous search of airborne release sources is intermittent sensor reading. Due to ultra low concentration, limited resolution of onboard sensors and local turbulence (possibly caused by the movement of sensor platforms or local airflow), there is a quite significant rate of undetection events. 
	Because of sensor dropouts, the noises are no longer unbiased, and consequently the performance of proposed algorithms will be degraded. As mentioned in Remark~\ref{rem: 7}, the proposed multi-estimator approach is in fact a hybrid method that combines the advantages of both model-based and model-free techniques. Consequently, it is able to cope with nonlinear and non-Gaussian filter problems. The performance of the CL-DCEE algorithm is satisfactory under intermittent sensor dropouts as shown via the simulation results.

		\begin{table}
				{\color{black}
		\setlength\heavyrulewidth{0.25ex}
		\centering
		\caption{Time consumed by running different algorithms for $200$ trials with unknown environment and sensor dropouts.}
		\label{tab: 3}
		\resizebox{\linewidth}{!}{%
			\begin{tabular}{@{}cccccccc@{}}
				\toprule
				& \multicolumn{6}{c}{CL-DCEE}      & Entrotaxis \\ \midrule
				Estimators/Particles & 5 & 10 & 50 & 100 & 200 & 1,000 & 10,000      \\
				Time (second)        &19.7   & 21.3   & 27.6   &28.2    &35.5     & 53.6      &   2877.9
				\\ \bottomrule
			\end{tabular}%
}		}
	\end{table}

\begin{figure}
		\centering
		\includegraphics[width=1\linewidth]{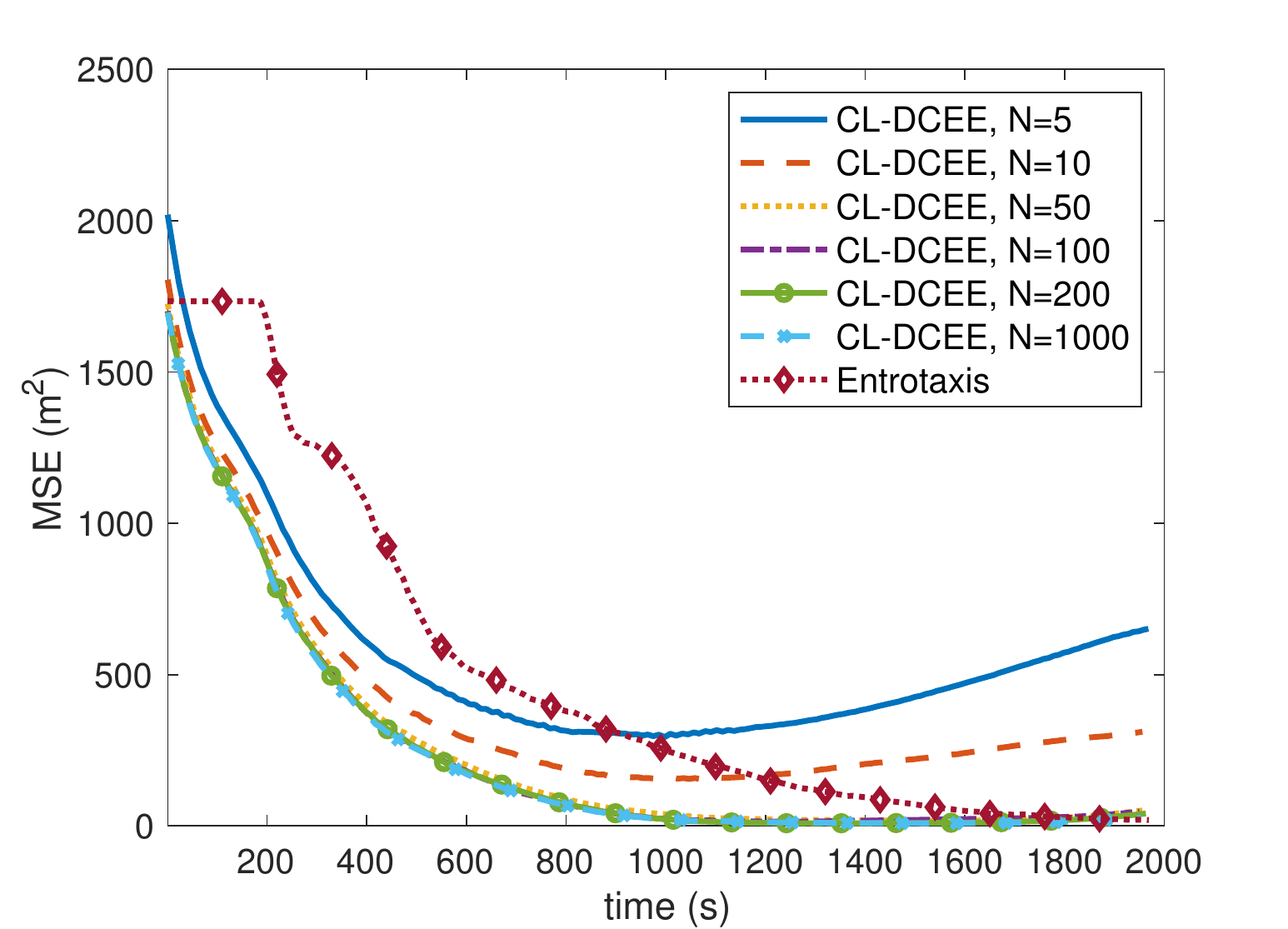}
		\caption{Mean-square-error between estimated and true source positions with unknown environment and sensor dropouts.}
		\label{fig: 4}
	\end{figure}

\begin{figure}
		\centering
		\includegraphics[width=1\linewidth]{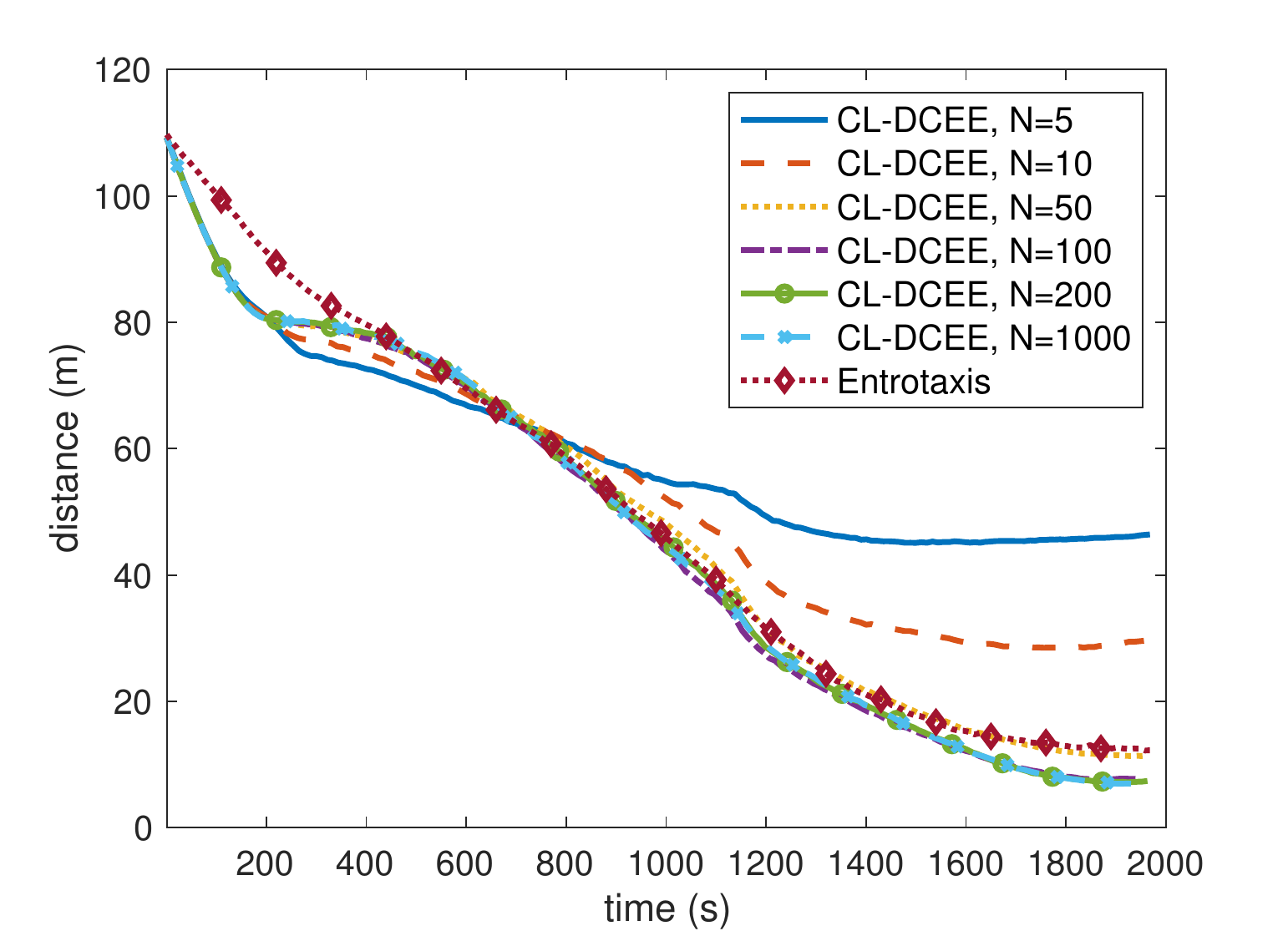}
		\caption{Distance between agent's position and the true source  with unknown environment and sensor dropouts.}
		\label{fig: 5}
	\end{figure}

	We have run each algorithm for $200$ times with random sensor dropouts. 
Table \ref{tab: 3} shows time consumed by running difference settings for $200$ trials. Although there are more uncertainties in the sensor and environment, the computational loads of those algorithms are similar. 
	Fig.~\ref{fig: 4} shows the MSE of source estimation.
	It is noticeable that the information-theoretic method exhibits quite strong resilience to the sensor dropouts and environment uncertainties. In Fig.~\ref{fig: 5}, distance between the agent and the true source is displayed.  Compared with previous results in Fig. \ref{fig: 2}, it is clear that the performance of our method has been influenced by the sensor dropouts and unknown environment, whereas Entrotaxis provides similar result as before. Nevertheless, all algorithms with different number of estimators can gradually navigate the agent move close to the source position. 
	Noticeably, the estimation performance of the CL-DCEE algorithm is degraded in the later period of search, in particular, for $N =5$ and $10$. This behaviour is probably due to the presence of non-detection events, which leads to biased measurement information. As the estimation process continues, such a biased measurement is gradually accumulated and worsens the search performance. The overall search performance and robustness to sensor dropouts are significantly enhanced by increasing the number of estimators.

	By the simulation comparison, it can be concluded that the information-theoretic algorithm is computationally intensive, but it demonstrates strong robustness to sensor dropouts and unknown environment. On the other hand, the proposed algorithm in this paper is much cheap to evaluate, but more sensitive to sensor performance and uncertainties. By comparing the obtained results in Sections \ref{sec: 4.1} and \ref{sec: 4.2}, we notice that knowing the environment can significantly enhance the performance of CL-DCEE algorithm, but has quite limited contribution for Entrotaxis. In general, our algorithm is expected to produce more accurate estimation under good sensor performance and less uncertain environment. When the number of unknown parameters in the model increases, it becomes more challenging in theory to satisfy the persistent excitation condition as discussed in Remarks~\ref{rem: 5} and \ref{rem: 6}.


\section{Experimental Study}\label{sec: 5}
	In this section, we test the proposed CL-DCEE algorithm using a real experimental dataset collected by COANDA Research \& Development Corporation using a large recirculating water channel \cite{Branko2016study}. Fluorescein dye was released at a constant rate from a narrow tube, and concentration data was collected over the entire search domain. More detailed descriptions of the experiment settings can be found in \cite{Branko2016study, Hutchinson2018Entrotaxis}. The dataset is composed of a total number of $340$ sequential frames, where each of them consists of $49\times98$ pixels. 
	
\begin{figure*}
		\centering
		\includegraphics[width=1\linewidth]{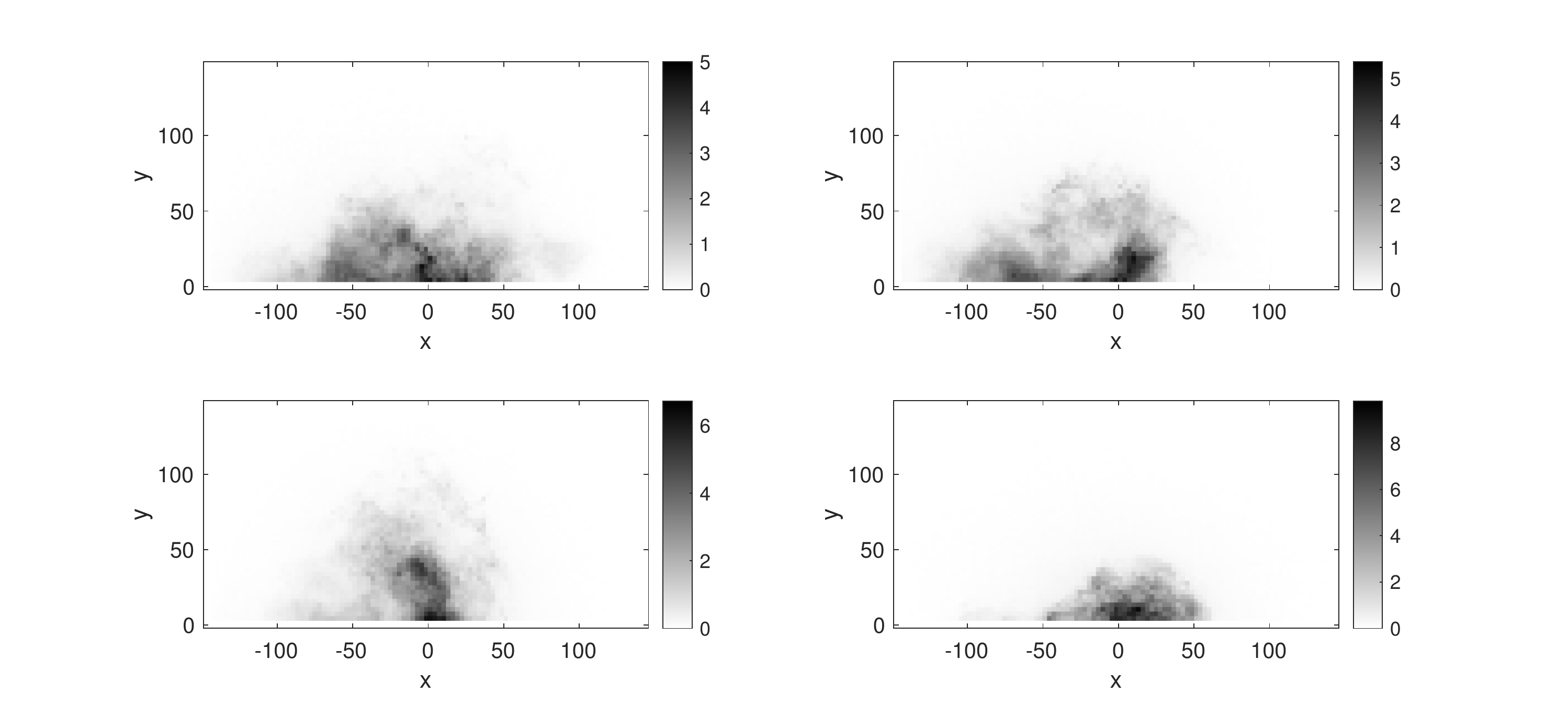}
		\caption{Concentration map of a dispersion field at different time frames where local turbulence changes the distribution dramatically from time to time. The sub-figures are taken at $1$, $2$, $3$ and $4$ sample instances, respectively. The grey-scale shade depicts the instantaneous concentration.}
		\label{fig: dispersion field only}
	\end{figure*}

    This is a quite challenging dataset for autonomous search due to rapid changes of the dispersion field. In Fig.~\ref{fig: dispersion field only}, four consecutive samples are depicted, which are collected at a sampling rate $10/23$s. During a very short period of time, the dispersion field changes very fast.  As a result, conventional reactive search approaches, like Chemotaxis, cannot obtain reliable sensor readings, and usually yield poor search performance. Estimation of the source parameters becomes essential in this type of challenging search scenarios. As discussed in Remark \ref{rem: 7}, the accuracy of model-based filtering techniques, like extended and unscented Kalman filter, are heavily influenced by the accuracy of the model and the assumed stochastic properties. In autonomous search, the models used in experiments, e.g., isotropic plume, may be quite different from the real dispersion field. Therefore, the hybrid ensemble strategy is beneficial for real-time computation of estimation variance since it inherits the advantages from both model-free and model-based techniques. 
		
\begin{figure*}
		\centering
		\includegraphics[width=1\linewidth]{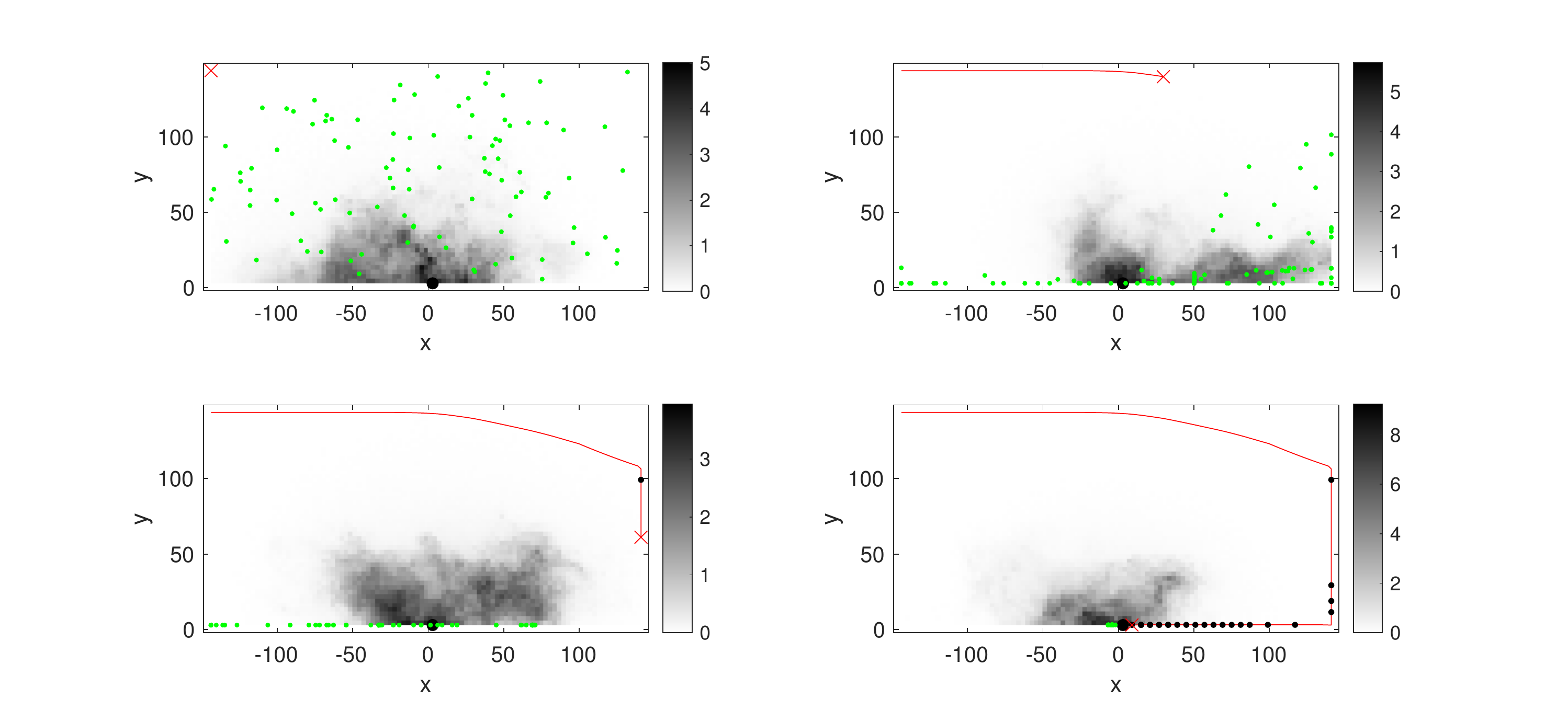}
		\caption{Representative search path of CL-DCEE on real dataset using an ensemble of $100$ estimators. The sub-figures are taken at $1$, $30$, $80$ and $146$ sample instances, respectively. Red lines are the paths of the search agent, red cross denotes agent's current position, the green dots represent the estimated source position, and the black dots represent non-zero measurements.}
		\label{fig: search path on real dataset using CL-DCEE}
	\end{figure*}

	Fig. \ref{fig: search path on real dataset using CL-DCEE} shows a representative search path using CL-DCEE algorithm. Most of the time, the sensor cannot receive any readings (i.e. undetected events) due to low concentration when the agent is far from the source. Until the $70$th iteration, the first meaningful reading is collected, as shown in the black dot in the third plot of Fig.~\ref{fig: search path on real dataset using CL-DCEE}. Fortunately, the estimators are able to update their estimates using zero measurements, because no measurement means that the source is less likely in the area. During the later period of searching, sensor can often obtain reliable readings as the concentration is high. Both estimators and the search agent converge to the source location, and the search mission is successfully completed as shown in the last plot of Fig. \ref{fig: search path on real dataset using CL-DCEE}.

	\section{Conclusion}\label{sec: 6}
	This paper has developed a computationally efficient solution for autonomous search of an airborne release with proven properties like convergence. A new learning framework, inspired by dual control for exploration and exploitation (DCEE), has been formulated to solve this goal-oriented control problem in an unknown environment with an unknown target.  Gradient-based optimisation algorithms have been proposed to estimate the source parameters, and to plan next movement by formulating suitable value functions. Theoretical guarantee for convergence and steady-state performance are analysed under measurement noises and uncertain turbulence. From the simulation and experimental studies, the effectiveness of the proposed solution has been validated. It has been demonstrated that our algorithm achieves superior performance comparing with informative path planning, and it also consumes much less computation time.

		One fundamental motivation for developing computationally efficient estimation approaches is to avoid prohibitive computational cost involved in predicting the future posteriors under a possible control action. In fact, the current setting in this paper is a myopic search strategy with one-step-ahead prediction. Previous work on multi-stage path planing is mainly based on certainty equivalence without considering future uncertainty, but it can provide better search trajectory by increasing the look ahead depth \cite{Bourgault2006optimal}. Active learning based non-myopic search has been recently pioneered in \cite{jiang2017efficient} for machine learning and data mining problems.  As discussed in \cite{Chen2021DCEE}, multi-stage dual control will be beneficial for enhancing search performance and robustness of autonomous search problem, but existing approaches are too computationally expensive to be implemented for multi-step DCEE. By employing the multi-estimator based approach developed in this paper, this will facilitate multi-stage DCEE so fully realise the potential of DCEE, which is a promising direction for future research.

	\bibliographystyle{IEEEtran}
	
	\bibliography{CLEEbib}

\begin{thebibliography}{10}
\providecommand{\url}[1]{#1}
\csname url@samestyle\endcsname
\providecommand{\newblock}{\relax}
\providecommand{\bibinfo}[2]{#2}
\providecommand{\BIBentrySTDinterwordspacing}{\spaceskip=0pt\relax}
\providecommand{\BIBentryALTinterwordstretchfactor}{4}
\providecommand{\BIBentryALTinterwordspacing}{\spaceskip=\fontdimen2\font plus
\BIBentryALTinterwordstretchfactor\fontdimen3\font minus
  \fontdimen4\font\relax}
\providecommand{\BIBforeignlanguage}[2]{{%
\expandafter\ifx\csname l@#1\endcsname\relax
\typeout{** WARNING: IEEEtran.bst: No hyphenation pattern has been}%
\typeout{** loaded for the language `#1'. Using the pattern for}%
\typeout{** the default language instead.}%
\else
\language=\csname l@#1\endcsname
\fi
#2}}
\providecommand{\BIBdecl}{\relax}
\BIBdecl

\bibitem{Hutchinson2017review}
M.~Hutchinson, H.~Oh, and W.-H. Chen, ``A review of source term estimation
  methods for atmospheric dispersion events using static or mobile sensors,''
  \emph{Information Fusion}, vol.~36, pp. 130--148, 2017.

\bibitem{Lennart1998source}
L.~Robertson and J.~Langner, ``Source function estimate by means of variational
  data assimilation applied to the {ETEX-I} tracer experiment,''
  \emph{Atmospheric Environment}, vol.~32, no.~24, pp. 4219--4225, 1998.

\bibitem{singh2015inverse}
S.~K. Singh, M.~Sharan, and J.-P. Issartel, ``Inverse modelling methods for
  identifying unknown releases in emergency scenarios: {An} overview,''
  \emph{International Journal of Environment and Pollution}, vol.~57, no. 1-2,
  pp. 68--91, 2015.

\bibitem{Rao2007source}
K.~Shankar~Rao, ``Source estimation methods for atmospheric dispersion,''
  \emph{Atmospheric Environment}, vol.~41, no.~33, pp. 6964--6973, 2007.

\bibitem{tsitsimpelis2019review}
I.~Tsitsimpelis, C.~J. Taylor, B.~Lennox, and M.~J. Joyce, ``A review of
  ground-based robotic systems for the characterization of nuclear
  environments,'' \emph{Progress in Nuclear Energy}, vol. 111, pp. 109--124,
  2019.

\bibitem{Hutchinson2018TCST}
M.~Hutchinson, C.~Liu, and W.-H. Chen, ``Information-based search for an
  atmospheric release using a mobile robot: Algorithm and experiments,''
  \emph{IEEE Transactions on Control Systems Technology}, vol.~27, no.~6, pp.
  2388--2402, 2018.

\bibitem{Chen2021DCEE}
W.-H. Chen, C.~Rhodes, and C.~Liu, ``Dual control for exploitation and
  exploration ({DCEE}) in autonomous search,'' \emph{Automatica}, vol. 133, no.
  109851, 2021.

\bibitem{vergassola2007nature}
M.~Vergassola, E.~Villermaux, and B.~I. Shraiman, ``Infotaxis as a strategy for
  searching without gradients,'' \emph{Nature}, vol. 445, no. 7126, pp.
  406--409, 2007.

\bibitem{pang2006chemical}
S.~Pang and J.~A. Farrell, ``Chemical plume source localization,'' \emph{IEEE
  Transactions on Systems, Man, and Cybernetics, Part B (Cybernetics)},
  vol.~36, no.~5, pp. 1068--1080, 2006.

\bibitem{Hutchinson2019magazine}
M.~Hutchinson, C.~Liu, P.~Thomas, and W.-H. Chen, ``Unmanned aerial
  vehicle-based hazardous materials response: Information-theoretic hazardous
  source search and reconstruction,'' \emph{IEEE Robotics \& Automation
  Magazine}, 2019.

\bibitem{Hutchinson2019fiedrobotics}
M.~Hutchinson, C.~Liu, and W.-H. Chen, ``Source term estimation of a hazardous
  airborne release using an unmanned aerial vehicle,'' \emph{Journal of Field
  Robotics}, vol.~36, no.~4, pp. 797--817, 2019.

\bibitem{yang2020optimal}
J.~Yang, C.~Liu, M.~Coombes, Y.~Yan, and W.-H. Chen, ``Optimal path following
  for small fixed-wing {UAVs} under wind disturbances,'' \emph{IEEE
  Transactions on Control Systems Technology}, vol.~29, no.~3, pp. 996--1008,
  2020.

\bibitem{chen2019odor}
X.-X. Chen and J.~Huang, ``Odor source localization algorithms on mobile
  robots: A review and future outlook,'' \emph{Robotics and Autonomous
  Systems}, vol. 112, pp. 123--136, 2019.

\bibitem{villa2016overview}
T.~F. Villa, F.~Gonzalez, B.~Miljievic, Z.~D. Ristovski, and L.~Morawska, ``An
  overview of small unmanned aerial vehicles for air quality measurements:
  Present applications and future prospectives,'' \emph{Sensors}, vol.~16,
  no.~7, p. 1072, 2016.

\bibitem{Hutchinson2018Entrotaxis}
M.~Hutchinson, H.~Oh, and W.-H. Chen, ``Entrotaxis as a strategy for autonomous
  search and source reconstruction in turbulent conditions,'' \emph{Information
  Fusion}, vol.~42, pp. 179--189, 2018.

\bibitem{Branko2016study}
B.~Ristic, A.~Skvortsov, and A.~Gunatilaka, ``A study of cognitive strategies
  for an autonomous search,'' \emph{Information Fusion}, vol.~28, pp. 1--9,
  2016.

\bibitem{zhao2020entrotaxis-jump}
Y.~Zhao, B.~Chen, Z.~Zhu, F.~Chen, Y.~Wang, and D.~Ma, ``Entrotaxis-jump as a
  hybrid search algorithm for seeking an unknown emission source in a
  large-scale area with road network constraint,'' \emph{Expert Systems with
  Applications}, vol. 157, p. 113484, 2020.

\bibitem{zhao2020searching}
Y.~Zhao, B.~Chen, Z.~Zhu, F.~Chen, Y.~Wang, and Y.~Ji, ``Searching the
  diffusive source in an unknown obstructed environment by cognitive strategies
  with forbidden areas,'' \emph{Building and Environment}, vol. 186, p. 107349,
  2020.

\bibitem{sun2011planning}
Y.~Sun, F.~Gomez, and J.~Schmidhuber, ``Planning to be surprised: Optimal
  bayesian exploration in dynamic environments,'' in \emph{International
  Conference on Artificial General Intelligence}.\hskip 1em plus 0.5em minus
  0.4em\relax Springer, 2011, Conference Proceedings, pp. 41--51.

\bibitem{liu2010stochastic}
S.-J. Liu and M.~Krstic, ``Stochastic source seeking for nonholonomic
  unicycle,'' \emph{Automatica}, vol.~46, no.~9, pp. 1443--1453, 2010.

\bibitem{Ramirez-Llanos2018stochastic}
E.~Ramirez-Llanos and S.~Martinez, ``Stochastic source seeking for mobile
  robots in obstacle environments via the {SPSA} method,'' \emph{IEEE
  Transactions on Automatic Control}, vol.~64, no.~4, pp. 1732--1739, 2018.

\bibitem{sapll1992multivariate}
J.~C. Spall, ``Multivariate stochastic approximation using a simultaneous
  perturbation gradient approximation,'' \emph{IEEE Transactions on Automatic
  Control}, vol.~37, no.~3, pp. 332--341, 1992.

\bibitem{ogren2004cooperative}
P.~Ogren, E.~Fiorelli, and N.~E. Leonard, ``Cooperative control of mobile
  sensor networks: Adaptive gradient climbing in a distributed environment,''
  \emph{IEEE Transactions on Automatic control}, vol.~49, no.~8, pp.
  1292--1302, 2004.

\bibitem{Bourgault2006optimal}
F.~Bourgault, T.~Furukawa, and H.~F. Durrant-Whyte, ``Optimal search for a lost
  target in a bayesian world,'' in \emph{Field and Service Robotics}.\hskip 1em
  plus 0.5em minus 0.4em\relax Springer, 2003, Conference Proceedings, pp.
  209--222.

\bibitem{li2011odor}
J.-G. Li, Q.-H. Meng, Y.~Wang, and M.~Zeng, ``Odor source localization using a
  mobile robot in outdoor airflow environments with a particle filter
  algorithm,'' \emph{Autonomous Robots}, vol.~30, no.~3, pp. 281--292, 2011.

\bibitem{ristic2003beyond}
B.~Ristic, S.~Arulampalam, and N.~Gordon, \emph{Beyond the Kalman filter:
  Particle Filters for Tracking Applications}.\hskip 1em plus 0.5em minus
  0.4em\relax Artech House, 2003.

\bibitem{Rhodes2021autonomous}
C.~Rhodes, C.~Liu, P.~Westoby, and W.-H. Chen, ``Autonomous search of an
  airborne release in urban environments using informed tree planning,''
  \emph{arXiv preprint arXiv:2109.03542}, 2021.

\bibitem{mesbah2018stochastic}
A.~Mesbah, ``Stochastic model predictive control with active uncertainty
  learning: A survey on dual control,'' \emph{Annual Reviews in Control},
  vol.~45, pp. 107--117, 2018.

\bibitem{chen2018approximating}
S.~Chen, K.~Saulnier, N.~Atanasov, D.~D. Lee, V.~Kumar, G.~J. Pappas, and
  M.~Morari, ``Approximating explicit model predictive control using
  constrained neural networks,'' in \emph{Annual American Control Conference
  (ACC)}.\hskip 1em plus 0.5em minus 0.4em\relax IEEE, 2018, Conference
  Proceedings, pp. 1520--1527.

\bibitem{bugeja2009dual}
M.~K. Bugeja, S.~G. Fabri, and L.~Camilleri, ``Dual adaptive dynamic control of
  mobile robots using neural networks,'' \emph{IEEE Transactions on Systems,
  Man, and Cybernetics, Part B (Cybernetics)}, vol.~39, no.~1, pp. 129--141,
  2008.

\bibitem{filatov2000survey}
N.~M. Filatov and H.~Unbehauen, ``Survey of adaptive dual control methods,''
  \emph{IEE Proceedings-Control Theory and Applications}, vol. 147, no.~1, pp.
  118--128, 2000.

\bibitem{li2011generalized}
S.~Li, J.~Yang, W.-H. Chen, and X.~Chen, ``Generalized extended state observer
  based control for systems with mismatched uncertainties,'' \emph{IEEE
  Transactions on Industrial Electronics}, vol.~59, no.~12, pp. 4792--4802,
  2011.

\bibitem{chen2015disturbance}
W.-H. Chen, J.~Yang, L.~Guo, and S.~Li, ``Disturbance-observer-based control
  and related methods: {An} overview,'' \emph{IEEE Transactions on Industrial
  Electronics}, vol.~63, no.~2, pp. 1083--1095, 2015.

\bibitem{goodwin1987parameter}
G.~C. Goodwin and D.~Q. Mayne, ``A parameter estimation perspective of
  continuous time model reference adaptive control,'' \emph{Automatica},
  vol.~23, no.~1, pp. 57--70, 1987.

\bibitem{holmes2006review}
N.~S. Holmes and L.~Morawska, ``A review of dispersion modelling and its
  application to the dispersion of particles: An overview of different
  dispersion models available,'' \emph{Atmospheric Environment}, vol.~40,
  no.~30, pp. 5902--5928, 2006.

\bibitem{stock2009chemotaxis}
J.~B. Stock and M.~Baker, ``Chemotaxis,'' in \emph{Encyclopedia of
  Microbiology}.\hskip 1em plus 0.5em minus 0.4em\relax Elsevier Inc., 2009,
  pp. 71--78.

\bibitem{welch1995introduction}
G.~Welch and G.~Bishop, \emph{An Introduction to the Kalman Filter}.\hskip 1em
  plus 0.5em minus 0.4em\relax Chapel Hill, NC, USA, 1995.

\bibitem{ortega2020modified}
R.~Ortega, V.~Nikiforov, and D.~Gerasimov, ``On modified parameter estimators
  for identification and adaptive control. a unified framework and some new
  schemes,'' \emph{Annual Reviews in Control}, 2020.

\bibitem{Dhariwal2004bacterium}
A.~Dhariwal, G.~S. Sukhatme, and A.~A. Requicha, ``Bacterium-inspired robots
  for environmental monitoring,'' in \emph{IEEE International Conference on
  Robotics and Automation}.\hskip 1em plus 0.5em minus 0.4em\relax IEEE, 2004,
  Conference Proceedings, pp. 1436--1443.

\bibitem{russell1995robotic}
R.~A. Russell, D.~Thiel, R.~Deveza, and A.~Mackay-Sim, ``A robotic system to
  locate hazardous chemical leaks,'' in \emph{Proceedings of IEEE International
  Conference on Robotics and Automation}.\hskip 1em plus 0.5em minus
  0.4em\relax IEEE, 1995, Conference Proceedings, pp. 556--561.

\bibitem{rudin1976principles}
W.~Rudin, \emph{Principles of Mathematical Analysis}, 3rd~ed.\hskip 1em plus
  0.5em minus 0.4em\relax New York, NY, USA: McGraw-hill, 1976.

\bibitem{guay2003adaptive}
M.~Guay and T.~Zhang, ``Adaptive extremum seeking control of nonlinear dynamic
  systems with parametric uncertainties,'' \emph{Automatica}, vol.~39, no.~7,
  pp. 1283--1293, 2003.

\bibitem{ding2007performance}
F.~Ding and T.~Chen, ``Performance analysis of multi-innovation gradient type
  identification methods,'' \emph{Automatica}, vol.~43, no.~1, pp. 1--14, 2007.

\bibitem{chua2018deep}
K.~Chua, R.~Calandra, R.~McAllister, and S.~Levine, ``Deep reinforcement
  learning in a handful of trials using probabilistic dynamics models,''
  \emph{arXiv preprint arXiv:1805.12114}, 2018.

\bibitem{Lakshminarayanan2017simple}
B.~Lakshminarayanan, A.~Pritzel, and C.~Blundell, ``Simple and scalable
  predictive uncertainty estimation using deep ensembles,'' \emph{Advances in
  Neural Information Processing Systems}, vol.~30, 2017.

\bibitem{chen2004disturbance}
W.-H. Chen, ``Disturbance observer based control for nonlinear systems,''
  \emph{IEEE/ASME Transactions on Mechatronics}, vol.~9, no.~4, pp. 706--710,
  2004.

\bibitem{barshalom1974dual}
Y.~Bar-Shalom and E.~Tse, ``Dual effect, certainty equivalence, and separation
  in stochastic control,'' \emph{IEEE Transactions on Automatic Control},
  vol.~19, no.~5, pp. 494--500, 1974.

\bibitem{fisac2018general}
J.~F. Fisac, A.~K. Akametalu, M.~N. Zeilinger, S.~Kaynama, J.~Gillula, and
  C.~J. Tomlin, ``A general safety framework for learning-based control in
  uncertain robotic systems,'' \emph{IEEE Transactions on Automatic Control},
  vol.~64, no.~7, pp. 2737--2752, 2018.

\bibitem{jeong2019learning}
H.~Jeong, B.~Schlotfeldt, H.~Hassani, M.~Morari, D.~D. Lee, and G.~J. Pappas,
  ``Learning {Q}-network for active information acquisition,'' \emph{arXiv
  preprint arXiv:1910.10754}, 2019.

\bibitem{guha2020MRAC}
A.~Guha and A.~Annaswamy, ``{MRAC-RL}: A framework for on-line policy
  adaptation under parametric model uncertainty,'' \emph{arXiv preprint
  arXiv:2011.10562}, 2020.

\bibitem{jiang2017efficient}
S.~Jiang, G.~Malkomes, G.~Converse, A.~Shofner, B.~Moseley, and R.~Garnett,
  ``Efficient nonmyopic active search,'' in \emph{International Conference on
  Machine Learning}.\hskip 1em plus 0.5em minus 0.4em\relax PMLR, 2017,
  Conference Proceedings, pp. 1714--1723.

\end{thebibliography}
	
	\begin{IEEEbiography}[{\includegraphics[width=1in,height=1.25in,clip,keepaspectratio]{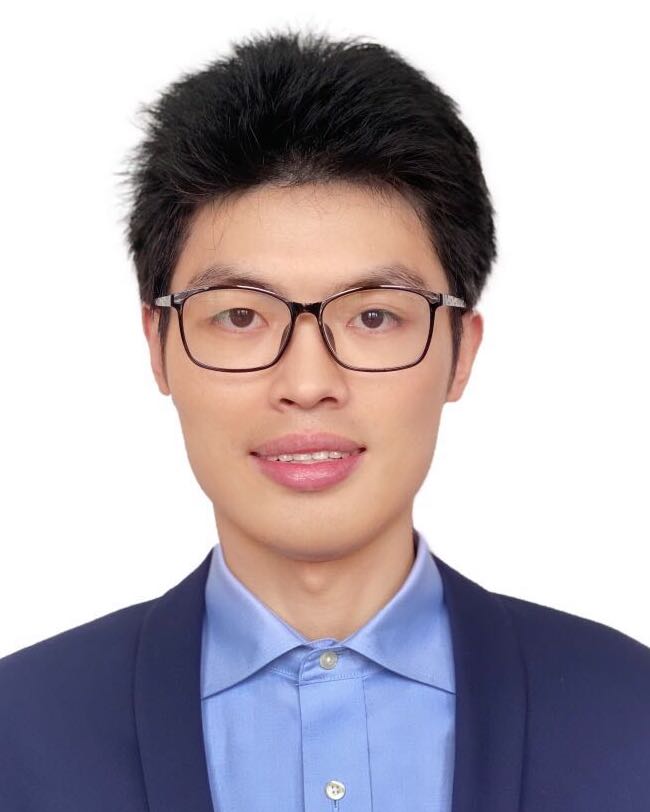}}]
		{Zhongguo Li} (Member, IEEE) received the B.Eng. and Ph.D. degrees in electrical and electronic engineering from the University of Manchester, Manchester, U.K., in 2017 and 2021, respectively.
		He is currently a Research Associate with the Department of Aeronautical and Automotive Engineering, Loughborough University, Loughborough, U.K., where he is involved in the EPSRC Established Career Fellowship Project ``Goal-Oriented Control Systems". His research interests include optimisation and decision-making for advanced control, distributed algorithm development for game and learning in network connected multi-agent systems, and their applications in autonomous vehicles.
	\end{IEEEbiography}

	\begin{IEEEbiography}[{\includegraphics[width=1in,height=1.25in,clip,keepaspectratio]{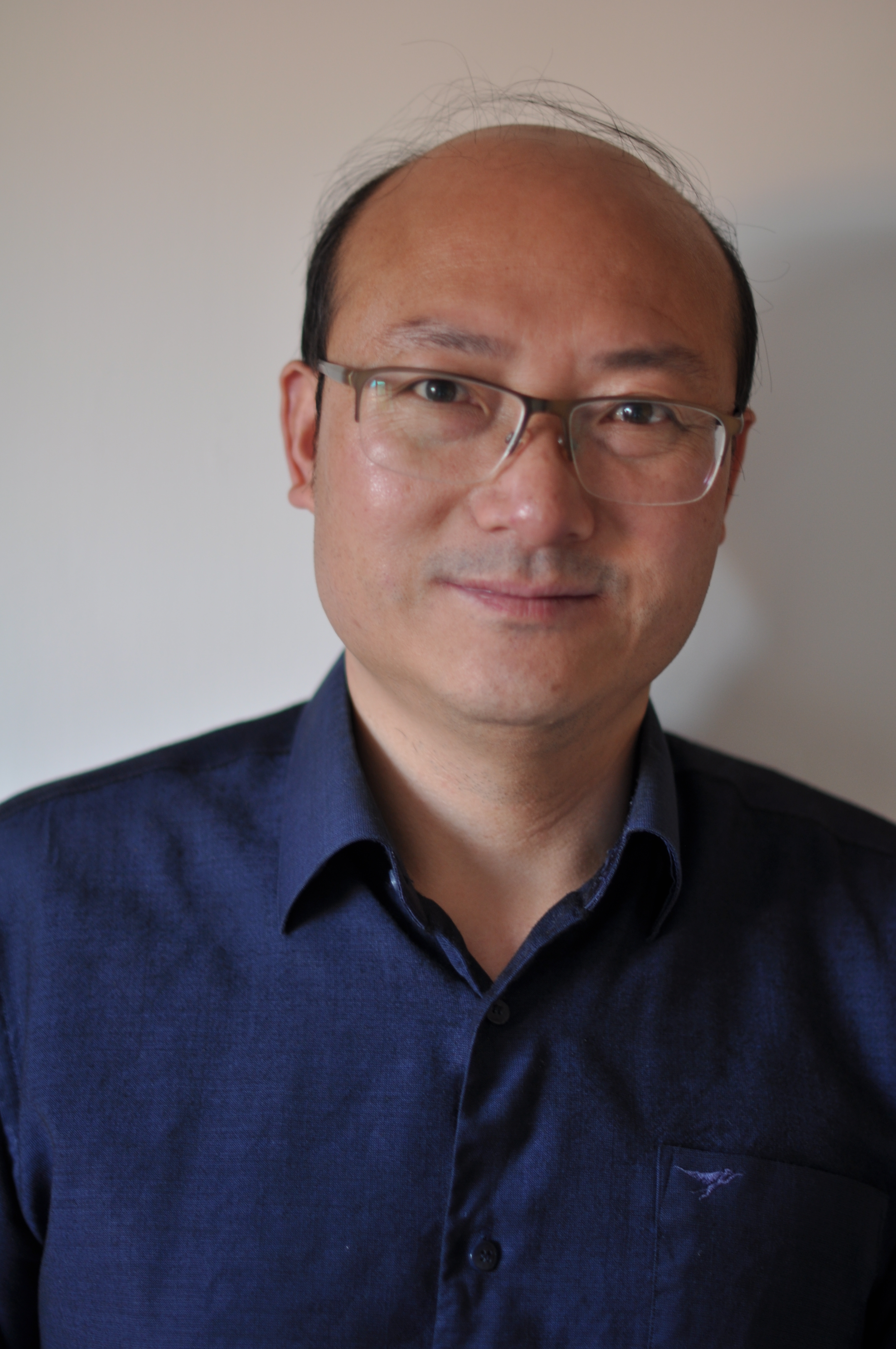}}]{Wen-Hua Chen} (M'00-SM'06-F'17)  holds a Chair in Autonomous Vehicles with the Department of Aeronautical and Automotive Engineering, Loughborough University, Loughborough, U.K., where he is leading the Centre of Autonomous Systems. He has a considerable experience in control, signal processing and artificial intelligence and their applications in robots, aerospace, and automotive systems. 
		He joined the Department of Aeronautical and Automotive Engineering, Loughborough University, in 2000 after having held a research position and then a Lecturer in control engineering with the Centre for Systems and Control, University of Glasgow, Scotland. Recently he was awarded a 5 years Established Career Fellowship by the UK Engineering and Physical Sciences Research Council. He is a Chartered Engineer, a Fellow of IEEE, the Institution of Mechanical Engineers and the Institution of Engineering and Technology, U.K. He has authored or coauthored near 300 papers and 2 books.
	\end{IEEEbiography}

	\begin{IEEEbiography}[{\includegraphics[width=1in,height=1.25in,clip,keepaspectratio]{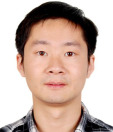}}]
		{Jun Yang} (Senior Member, IEEE) received the B.Sc. degree from the Department of Automatic Control, Northeastern University, Shenyang, China, in 2006, and the Ph.D. degree in control theory and control engineering from the School of Automation, Southeast University, Nanjing, China, in 2011. He is currently a Senior Lecturer with the Department of Aeronautical and Automotive Engineering at Loughborough University. He is an IET Fellow. He received the ICI prize for best paper of Transactions of the Institute of Measurement and Control in 2016 and Premium Award for best paper of IET Control Theory and Applications in 2017. His current research interests include disturbance estimation and compensation, advanced control theory, and its application to electric machines, mechatronic systems and robotics.
	\end{IEEEbiography}
	
\end{document}